\documentclass[twocolumn,american,nofootinbib]{revtex4-2}
\usepackage{amsmath}
\usepackage{newtxmath}
\usepackage[T1]{fontenc}
\usepackage{textcomp}
\usepackage[utf8]{inputenc}
\usepackage{xcolor}
\usepackage{babel}
\usepackage{booktabs}
\usepackage{graphicx}
\usepackage[pdfusetitle,
 bookmarks=true,bookmarksnumbered=false,bookmarksopen=false,
 breaklinks=false,pdfborder={0 0 1},backref=false,colorlinks=true]
 {hyperref}
\hypersetup{
 urlcolor=blue, citecolor=blue, linkcolor=blue}

\makeatletter

\usepackage{float}
\usepackage{xcolor}
\usepackage{soul}
\usepackage[normalem]{ulem}

\makeatother

\begin{document}

\title[]{On Phase Transition of ITG Turbulence in the Dimits shift}

\author{L. N. Marquant}

\author{P. Morel}
 
\author{Ö. D. Gürcan}

\affiliation{Laboratoire de Physique des Plasmas, CNRS, Ecole Polytechnique, Sorbonne Université, Université Paris-Saclay
, Observatoire de Paris, F-91120 Palaiseau, France}


\begin{abstract}
The transition between turbulent and zonal flow dominated states is investigated by varying the ion temperature gradient in nonlinear gyrokinetic simulations.
Independent gradient scans reveal three distinct regimes: a turbulent regime at high gradients,
a zonal flow dominated regime with strongly reduced heat transport at low gradients, and an intermediate regime characterized by intermittent switching between these two states.
These different regimes can be classified using an order parameter, defined as the fraction of the zonal to total free energy in the system. 
To assess the memory effects, the temperature gradient is first lowered gradually from high values that result in fully developed turbulence,
to lower values in the Dimits shift region that form strong zonal flows, and then is slowly increased back. Once the zonal flows form, and efficiently suppress turbulence,
they can persist at higher gradients, leading to an asymmetric response implying a hysteresis loop. It is observed that, in the zonal flow dominated state,
the free energy is mostly condensated in the largest radial scale, with a steep slope of the $k_x$ spectrum, while in the turbulent state, it exhibits a wider spectrum with two distinct slopes.

\end{abstract}

\maketitle

\section{\label{sec:level1}Introduction}

Understanding micro-turbulence is crucial for achieving controlled nuclear fusion in magnetized plasmas as it limits the plasma confinement time through \emph{anomalous transport} \citep{Groebner1986, Carreras1997}
driven by turbulent fluctuations \citep{Wootton1990}.
Among the various types of instabilities possible in tokamaks, ion temperature gradient (ITG) \citep{Coppi1967} driven modes are considered one of the dominant mechanisms responsible for the ion channel turbulent transport.
ITG driven turbulence develops at the ion Larmor radius (i.e. $\rho_i$) scales, and lies primarily in the plane perpendicular to the magnetic field.
It exhibits a linear threshold for the ion temperature gradient, which can be expressed in terms of the parameter $R/L_{T_i} > R/L_{T_i,c}$,
where $R$ is the major radius of the tokamak and $L_{T_i} = |d \log T_i / dr|^{-1}$ is the gradient length.
ITG also presents a nonlinear threshold, because the near marginal turbulence that it produces self-regulates by generating axisymmetric poloidal structures called zonal flows which
can suppress the turbulence that drive them. 

The Dimits shift \citep{Dimits2000} is commonly characterized as a nonlinear shift of the critical gradient needed to obtain significant anomalous heat diffusivity.
The details of how this shift is observed in numerical simulations depends on various parameters,
such as magnetic geometry, numerical resolution  \citep{Peeters2016} or collisionality \citep{Weikl2017}.
One key observation that may shed light into the nature of this shifted threshold is that,
for moderate temperature gradients, associated to quasi-null radial transport, the time traces of heat flux are found to be alternating intermittently between non-zero, and quasi-null, plateau-like values of the heat flux.
This suggest that the Dimits shift could be interpreted as being analogous to a "phase transition" between a zonal dominated phase at moderately small temperature gradients just above the linear stability threshold,
and a turbulence dominated phase for large enough temperature gradients.

Notice that, the Low to High confinement (L-H) transition \citep{Connor2000}, observed in most tokamaks \citep{WagnerASDEXLH, Burrell1992}, can also be interpreted as an hysteretic phase transition \citep{Malkov2009},
a generalization to hysteresis in flux-gradient relationship in magnetized plasma has been proposed for experimental data analysis as well \citep{Itoh2017}. 

Recently, the transition between the two dimensional turbulence and the zonal flow dominated state in the Hasegawa-Wakatani model of drift-wave turbulence is demonstrated to constitute
an almost textbook case of a thermodynamical phase transition, with a clear hysteresis loop,
where the ratio of adiabaticity to the background gradient defines the control parameter and the ratio of zonal to total kinetic energy defines the order parameter.
The mechanism of the hysteresis is that once the zonal flows are formed, it takes more free energy to break them apart in loose analogy with \emph{latent heat} in crystal melting.
Earlier results showing similar hysteresis behavior, were observed in trapped particle turbulence simulations  \citep{Gravier2017},
and in the Hasegawa-Wakatani system with finite Larmor radius corrections \citep{Grander2024, Guillon2025}.
Recent dicussion of whether or not this is an actual phase transition or simply a subcritical zonal bifurcation which resembles a phase transition is also probably relevant  \citep{grander2026mergingzonalflowsgyrofluid}.

In the present work, similar results are recovered using electrostatic gyrokinetic simulations with adiabatic electrons. 
Adopting the normalized ion temperature gradient as the control parameter and the zonal free energy fraction as the order parameter, 
the Dimits nonlinear upshift is shown to be associated to a phase transition between zonal and turbulence dominated regimes.
In the local approach adopted here, i.e. gradient-driven micro-turbulence, such a transition appears robust, and is examined into details for two scenarios: 
independent simulations for various temperature gradients, and chained simulations obtained by restarting the same simulation while progressively varying the temperature gradient.
An hysteresis can possibly be present, and two main parameters govern its appearance: the rate of change of the temperature gradient and the zonal flow dissipation.
A lower rate of change (i.e. long simulation per $R/L_{T,i}$) increases the likelihood of observing an hysteresis,, with a nearly systematic occurrence at sufficiently low rates,
whereas higher rates reduce the probability of finding an hysteretic behavior. Similarly, reducing the zonal flow dissipation also favors the appearance of hysteresis.

The paper is organized as follows: in section \ref{sec:GENE} the local gyrokinetic equation that is solved by the GENE code is presented together with the typical parameters that are used.
In the next section \ref{sec:DimitsShift}, the basic physics of the Dimits shift is reminded. In section \ref{sec:ramps}, particular emphasis is placed on chained simulations,
in which the ion temperature gradient is gradually varied in both decreasing and increasing directions to explore the system’s behavior across the threshold.
From a physical perspective, this procedure can be interpreted as playing on the heating. Before concluding remarks (section \ref{sec:Conclusion}),
the influence of varying the ion temperature gradient on the turbulent spectra is compared between chained and independent series of simulations in section \ref{sec:spectra}.

\section{Non-linear simulation: Numerical Experiment \label{sec:GENE}}

All simulations presented in this study are performed using the electrostatic version of the gyrokinetic code GENE \citep{Jenko2000} using a local flux-tube formulation.
GENE is an Eulerian code, which solves the nonlinear electromagnetic gyrokinetic equations \citep{Brizard2007} on a five-dimensional fixed grid in phase space for an arbitrary number of species,
in a given magnetic equilibrium. It uses a field-aligned coordinate system  $(x,y,z,v_{\parallel},\mu)$ in order to take advantage of the strong anisotropy of plasma turbulence  \citep{Beer1994},
characterized by large correlation lengths along the magnetic field lines and much smaller correlation lengths in the perpendicular plane.

The present work studies the Dimits shift, which has been observed in ITG driven turbulence, while varying the ion temperature gradient, $R/L_{T,i}$, around ITG marginal stability,
for a set of parameters associated to a DIII-D shot, usually referred to as "the Cyclone Base Case", detailed in the following subsection \ref{subsec:CBC}.
The gyrokinetic equations solved in the local, gradient-driven, electrostatic version of the GENE code are presented \ref{subsec:GENE},
before discussing the zonal dissipation, which is crucial for our study \ref{subsec:ZonalDiss}.
The section ends with defining the two observables that will be used for analysis in the rest of the paper: free energy and ion heat flux \ref{subsec:FreeEnergy}.

In the following, simulations in which $R/L_{T,i}$ is held constant throughout the duration of the run will be referred to as \textit{independent} simulations.
In contrast, simulations where $R/L_{T,i}$ is discretely and slowly varied in time will be referred to as \textit{chained simulations},
since they consist of simulations that are stitched back to back where one uses the final state of one as the initial condition for the next.

\subsection{Standard parameters: Cyclone Base Case \label{subsec:CBC}}

\begin{table*}
\caption{Simulation grid, domain, and dissipation parameters for different cases.}
\label{tab:simulation_params}
\begin{ruledtabular}
\begin{tabular}{cccccc|cccc|cccc|c}
Case & $N_x$ & $N_{k_y}$ & $N_z$ & $N_{v_{||}}$ & $N_{\mu}$ & 
$l_x$ & $k_{y,\min}$ & $l_v$ & $l_\mu$ & 
\multicolumn{4}{c|}{Dissipation coefficients $(c)$} & Notes \\ 
\cline{11-14}
 & & & & & & & & & & $c_x$ & $c_y$ & $c_z$ & $c_{v_{\parallel}}$ & \\ \hline
\texttt{CBC}       & 96  & 16 & 16 & 32 & 8 & 125.628 & 0.05 & 3.00 & 9.00 & LES & LES & 2.0 & 0.20 & Baseline CBC resolution \\
\texttt{MIN\_C}   & 96  & 16 & 16 & 32 & 8 & 125.628 & 0.05 & 3.00 & 9.00 & 0     & 0.10  & 2.0 & 0.01 & CBC with minimized ZF dissipation \\
\end{tabular}
\end{ruledtabular}
\end{table*}

As mentioned above, apart from the normalized temperature gradient $R/L_{T,i}$, which is varied,  all the parameters are fixed at their corresponding \textit{Cyclone Base Case} (CBC) values,
namely: magnetic shear $\hat{s} = (r/q)dq/dr = 0.796$, inverse aspect ratio $\epsilon = r/R = 0.18$, safety factor $q = rB_t/(RB_p) = 1.4$  assuming circular magnetic flux surfaces, for simplicity.
The duration for each temperature gradient step is set to $N \times \gamma_{\max}^{-1}\!\left(R/L_{T,i}\right)$ where $\gamma$ denotes the linear growth rate computed separately using the linear diagnostic of GENE.
An appropriate value for $N$ in this study is at least $N\ge 1000$.
The simulation domain size and the resolution of the computational grid are given in Table~\ref{tab:simulation_params}.

\subsection{Gyrokinetic equations solved in GENE \label{subsec:GENE}}

For this study, we consider electrostatic ITG turbulence, where the plasma is assumed to be collisionless, consisting of a single ion species, with adiabatic electrons.
This is the simplest model of gyrokinetic ITG, with no particle transport since the electrons are adiabatic.
This means, only the ion guiding center distribution function, coupled to the electrostatic potential through the Poisson equation, needs to be solved.
Additionally, a $\delta f$ approach is employed, in which the distribution function is decomposed into a fixed background equilibrium component and an unknown perturbation,
$F_i = F_{0,i} + f_i$, with the equilibrium distribution $F_{0,i}$ taken to be Maxwellian, with temperature and density gradients.
The perturbation $f_i$ is then taken to be a function of the guiding center coordinates and evolved in the five-dimensional phase space $(k_x, k_y, z, v_{\parallel}, \mu)$,
where the perpendicular directions are treated in Fourier space, taking advantage of the periodic boundary conditions inherent to the flux-tube formulation.

After normalization, the governing equations in GENE take the following form:

\begin{align}
  \label{eq:Vlasov_gene}
  \partial_t f_{i,k} &= 
      \mathcal{L}_G\!\left[ J_{0k}\phi_k \right] 
    + \mathcal{L}_{C}\!\left[ f_{i,k} \right] 
    + \mathcal{L}_{\parallel}\!\left[ f_{i,k} \right] \nonumber \\ 
  &\quad + \mathcal{N}\!\left[ f_{i,k}, J_{0k}\phi_k \right] 
    + \mathcal{D}\!\left[ f_{i,k} \right], \\[0.9em]
  \label{eq:Poisson_gene}
  \phi_k - \langle \phi_k \rangle_{FS} 
  &+ \frac{Z_i T_{e0}}{T_{i0}} \, \big[ 1 - \Gamma_0(b_i) \big] \, \phi_k \nonumber \\
  &= \pi B_0 \int \! \mathrm{d}v_{\parallel} \, \mathrm{d}\mu \, J_{0k} f_{i,k}.
\end{align}

The quantities $f_{i,k}$ and $\phi_{k}$ denote, respectively, the perturbed ion guiding center distribution function and the perturbed electrostatic potential in Fourier Space.
$h_{i,k}$ refers to the non-adiabatic part of the ion guiding center distribution function and is defined as $h_{ik}= f_{ik} + q_i F_{0,i} J_{0k}\phi_k/T_{0,i}$,
 where $F_{0,i}$ is the equilibrium Maxwellian distribution, $q_i = Z_i e$ is the ion charge, $T_{0,i}$ is the background temperature
and $J_{0k}\phi_k$ denotes the gyroaveraged electrostatic potential with $J_{0k}$ being the zeroth-order Bessel function of the first kind.
The bracket operator $\langle X \rangle_{FS}= \int ( J\, X \, dz ) / ( \int J \, dz )$ indicates the flux surface average,
while  $\Gamma_0(b_i)$ denotes the modified bessel function of the first kind with argument $b_i=v^2_{T,i}k^2_{\perp}/\Omega^2_{ci}$
where $v_{T,i}\,,\Omega_{c,i}$ correspond to the ion thermal speed and the ion cyclotron frequency, respectively.

$\mathcal{L}_X$ represents linear terms capturing specific physical effects. $\mathcal{L}_G[J_{0k}\phi_k] = -(R/L_{n,i} + (v_\parallel^2 + \mu B_0 - 3/2)\, R/L_{T,i}) F_{0,i}\, i k_y J_{0k}\phi_k$
represents the drive associated with fixed background ion density and temperature gradients.
This is the only term that depends explicitly on $R/L_{T,i}$. The magnetic curvature effect is accounted in $\mathcal{L}_C[f_{ik}]$, while
$\mathcal{L}_{\parallel}[f_{ik}]$ correspond to the parallel dynamics, including both trapped-particle effects and Landau damping.

The term $\mathcal{N}[f_{i,k} , J_{0k} \phi_k ]$ denotes the nonlinear operator arising from the nonlinear advection of the perturbed ion guiding center distribution function
by the $E \times B$ drift velocity associated with the gyroaveraged electrostatic potential $J_{0k} \phi_k$.
It is the only nonlinearity retained in the equations, since the parallel nonlinearity \citep{Candy2006} is an order $\mathcal{O}(\rho^{\star} = \rho_L / R_0)$ higher than the other terms.

$\mathcal{D}$ is a numerical dissipation operator applied to the 4 phase-space coordinate $(k_x, k_y, z, v_{\parallel})$. It can be decomposed into:
a fourth-order hyperdiffusion operator ensuring parallel dissipation $\mathcal{D}_{\parallel}=-(c_z \partial^4_z
+ c_{v_{||}} \partial^4_{v_{||}})$, with $c_z,c_{v_{\parallel}}$ held constant, and 
a perpendicular dissipation operator $\mathcal{D}_{\perp}=- (c_x k^4_x + c_y k^4_y)$  which acts on the radial and binormal directions. 
In GENE, by default, the perpendicular dissipation coefficients are handled using the Large Eddy Simulation method \cite{Morel2011}.
LES is a numerical technique in which the unresolved small scales are modeled by dissipation operators, while the dissipation coefficients $c_x$ and $c_y$ are dynamically optimized during the simulation \cite{Morel2012}.

Since the objective of this study is to investigate into details the role of zonal flows, fixed dissipation coefficients $c_x$ and $c_y$ will be preferred to LES, in order to better tune the zonal and nonzonal dissipations.

If both parallel and perpendicular dissipation are set to a fourth-order hyperdiffusion operator,
the full dissipation operator is described by $\mathcal{D} = - (c_x k^4_x + c_y k^4_y + c_z \partial^4_z + c_{v_{||}} \partial^4_{v_{\parallel}})$ where $(c_x,c_y,c_z,c_{v_{\parallel}})$ are held constant.

\subsection{Zonal dissipation \label{subsec:ZonalDiss}}

\begin{figure}[H]
    \centering
    \includegraphics[width=\linewidth]{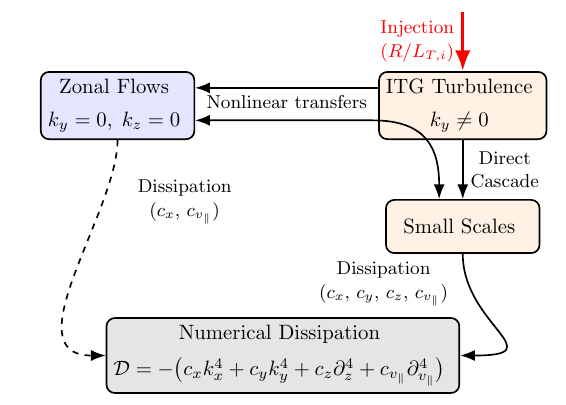}
    \caption{Schematic representation of energy transfers between turbulence, zonal flows, and numerical dissipation.}
    \label{fig:Fig_1}
\end{figure}

The distribution of free energy within the system can be understood through the schematic representation shown in Figure~\ref{fig:Fig_1},  which illustrates the nonlinear interactions between turbulence, zonal flows,
and numerical dissipation.

In ion-temperature-gradient (ITG) turbulence, energy is injected linearly when the radial temperature gradient exceeds a critical threshold \citep{Romanelli1989}.
This free energy is then redistributed across different spatial scales through nonlinear interactions. 

In contrast, zonal flows (ZFs) do not receive free energy through linear processes \citep{Diamond2005Review, Gurcan2015}, since ZFs are constant over a flux surface,
i.e. constant poloidally and toroidally. Consequently, zonal flows acquire their energy exclusively from turbulence via nonlinear interactions,
effectively acting as an energy sink that depletes turbulent energy.
Additionally, turbulence can be suppressed by flow shear decorrelation: after the linear growth of ITG instability,
nonlinear mode coupling transfers energy into zonal flows which can lead to the formation of radially sheared $E\times B$ flow structures.
This radial shear originated from zonal flows distorts turbulent eddies, increasing their radial wavenumber and reducing their correlation time.
As a result, eddies decorrelate faster, they have less time to get amplified and therefore turbulence and transport is diminished.
Moreover, if the shearing rate is sufficiently strong, it can even overcome the linear drive, changing the underlying linear physics of the system as long as these structures are present \citep{Biglari1990}.

This feature constitutes the main mechanism of turbulence suppression by zonal flow shear, which explains why ZFs are of major importance and have been extensively studied.

The complex dynamics between Zonal flows and turbulence can be interpreted via predator prey analogy \citep{Malkov2001, Schmitz2012, Morel2014, Kobayashi2015},
ZFs acting as a sink for turbulence by feeding on turbulent energy via nonlinear interaction and also decreasing the turbulence correlation length through radial flow shear.
Turbulence can therefore be seen as a prey, predated by ZFs.
In this analogy, one of the key parameter is the damping of the predator population as it describe how long the predator population (ZFs) will sustain when the prey population (turbulence) is low.

One of the main mechanism that depletes zonal flow energy is collisions. In the presence of collisional damping,
the turbulence suppression by zonal flow shear exhibits a finite lifetime, after which turbulence can reemerge while zonal flows asymptotically disappear \citep{Lin1999}.
In the collisionless limit, it has been shown that zonal flows converge to a finite, non-zero level \citep{Rosenbluth1998}.
In the framework of the present study, the system is not completely collisionless,
due to the use of the aforediscussed dissipation operator $\mathcal{D}$: since the subject of this study is the transition to a zonal flow dominated state, strong and persistent ZFs should be sustained,
so that the limit of low zonal dissipation is obtained by tuning the coefficients $(c_x,\, c_y,\, c_z,\, c_{v_\parallel})$ in order to minimize collisional effects.

Due to their geometry, zonal flows are unaffected by the dissipation in parallel and/or binormal directions.
Consequently, the terms proportional to $c_yk^4_y$ and $c_z\partial^4_z$ do not contribute to zonal flow damping.
Zonal flow damping comes only from the radial diffusion operator (which provides small scale dissipation), and the diffusion in parallel velocity.

Imposing $c_x = 0$, $c_y = 0.1$, $c_z = 2.00$, and $c_{v_\parallel} = 0.01$ leads to the minimized set-up referred to as \texttt{MIN\_C} in Table~\ref{tab:simulation_params}.
The dissipation along the parallel velocity direction is maintained non-zero to prevent recurrence phenomena \citep{Pueschel2010},
which result from the discrete representation of the velocity-space grid and cause the artificial reappearance of damped modes after long simulation times.

\subsection{Free energy / Heat Flux \label{subsec:FreeEnergy}}

Here we consider the evolution of free energy, a quadratic and positive-definite quantity conserved nonlinearly by the gyrokinetic equation \citep{Schekochihin2008}, which is defined as

\begin{equation}
\mathcal{E} = \int d\Lambda\, \frac{T_{0,i}}{F_{0,i}} \frac{h_i^2}{2},
\end{equation} where the phase-space integral \( d\Lambda \) is given by

\begin{equation}
\int d\Lambda = \int dx \int dy \int dz \int dv_{||} \int d\mu\, \pi B_0 n_{0,i},
\end{equation}

The time evolution of the free energy can be obtained by applying the operator $\mathfrak{X}[\cdot]$:

\begin{equation}
\mathfrak{X}[\cdot] = \frac{n_{0i} T_{0i}}{V T_{0e}} 
\sum_{k_x, k_y} 
\int \pi \, \mathrm{d}z \, \mathrm{d}v_{\parallel} \, \mathrm{d}\mu \, 
\frac{h_{i,k}^{*}}{F_{0i}} \, ,
\end{equation} to the Vlasov equation \eqref{eq:Vlasov_gene}, and then combining the result with the Poisson equation \eqref{eq:Poisson_gene}, yielding:

\begin{equation}
  \label{eq: Free_Energy_Production}
  \partial_t \mathcal{E} = {\mathcal{G}} - \mathcal{D} \, ,
\end{equation} where $\mathcal{G}$ represents injection, while $\mathcal{D}$ denotes dissipation of free energy.
It should be noted that the parallel and magnetic curvature effects do not contribute to the global free energy balance \citep{Banon2011}.
More importantly, since it is a nonlinearly conserved quantity, there is no net injection or dissipation of free energy by nonlinear interactions, but only redistribution between turbulent scales.

It is also useful to distinguish two subsets of quantities, namely the zonal and turbulent contributions to the free energy:

\begin{eqnarray}
  \label{eq: zonal free energy contribution}
  \overline{\mathcal{E}}_{k_x}(t) & \equiv & \mathcal{E}_{k_x,k_y=0}(t), 
  \quad \quad \text{(zonal contribution)} \, ,\\
  \widetilde{\mathcal{E}}_{k_x}(t) & \equiv & 2 \sum_{k_y > 0} \mathcal{E}_{{k_x,k_y}}(t),   \quad \text{(turbulent contribution)} \, .
  \label{eq: turbulent free energy contribution}
\end{eqnarray}

The factor 2 accounts for the contribution of the negative $k_y$ modes, which are omitted in the GENE output due to Hermitian symmetry of the free energy.
The zonal modes on the other hand, have $k_y=0$ and are therefore self conjugate and hence shouldn't be counted twice.
The total, time evolving, free energy contributions associated to the zonal and turbulent components are finally obtained by summing over all $k_x$ modes:

\begin{eqnarray}
  \label{eq: time zonal free energy contribution}
  \overline{\mathcal{E}}(t) & = & \sum_{k_x} \overline{\mathcal{E}}_{k_x}(t) \, , \\
  \widetilde{\mathcal{E}}(t) & = & \sum_{k_x} \widetilde{\mathcal{E}}_{k_x}(t) \, .
  \label{eq: time turbulent free energy contribution}
\end{eqnarray}

Normalizing these quantities by the total free energy yields the zonal and turbulent free energy fractions of the system respectively.
Using the zonal free energy fraction $\Xi \equiv \bar{\mathcal{E}}/\mathcal{E}$ as the order parameter allows us to characterize how much the system can be dominated by the zonal flows: when $\Xi \rightarrow 0$,
the total free energy is mainly due to turbulent contributions,  whereas $\Xi \rightarrow 1$ indicates that the system is dominated by zonal flows.

In addition to the order parameter $\Xi$, we consider the ion heat flux, which is the usual quantity associated to ITG turbulence and the Dimits shift, and can be expressed in our notation as:

\begin{equation}
Q = \int d^3v\, \frac{m_i v^2}{2} f_1\, v_{d,r} \, ,
\end{equation} 

where $v_{d,r}$ denotes the radial component of the generalized $\mathbf{E} \times \mathbf{B}$ drift velocity.
It is possible to relate the heat flux directly to the free energy injection term $\mathcal{G}$ through the relation:

\begin{equation}
\mathcal{G} = \frac{R}{L_{T,i}}\, Q.
\end{equation}

\section{Zonal flow damping, and the Dimits shift \label{sec:DimitsShift}}

In this section, independent simulations are performed around the Dimits threshold in order to clarify the effect of zonal flow damping.
First, a brief comparison between the baseline \texttt{CBC} model and the simulations 
with minimized zonal damping (\texttt{MIN\_C}) is carried out using free energy diagnostics. 
Here, in accordance with its original definition, we use the ion heat flux in order to characterize where the Dimits shift happens.
For this purpose, simulations are performed in the range $R/L_{T,i} = [5.0, 7.5]$, making sure that the zonal flows reach saturation, which is important to distinguish whether or not we are in the Dimits regime.
Accordingly, the following simulations are integrated for a long time to ensure that the system reaches a quasi-steady state.
As a reminder, the duration of each simulation is set by $N \times \gamma^{-1}_{\max}(R/L_{T,i})$,
where $\gamma^{-1}_{\max}$ denotes the maximum linear growth rate for a given temperature gradient, 
and the parameter $N$ is usually a large integer (typically in the thousands) chosen to ensure convergence.

\subsection{Validation of the simulation parameters at boundaries}

First, the two extreme cases of the temperature gradient are compared between the two sets of parameters \ref{tab:simulation_params}. 
Each simulation was run for $N=2000$, i.e. $2000 \, \gamma^{-1}_{\max} \left ( R/L_{T,i} \right )$, in order to investigate convergence and compare the two numerical set-ups. 

\begin{figure}[hbtp]
  \includegraphics[width=\linewidth]{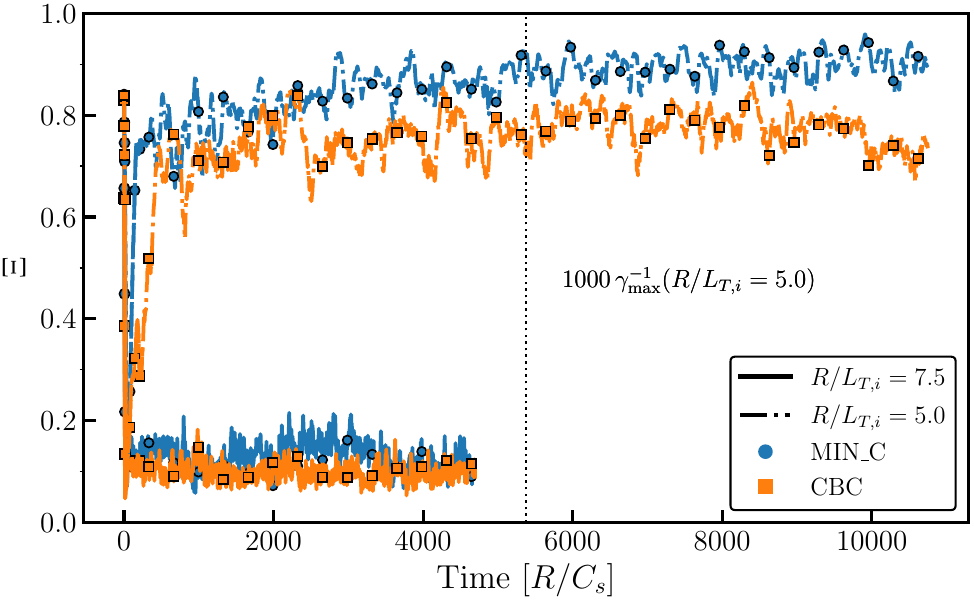}
  \caption{Time evolution of the zonal free energy fraction $\Xi$ for \texttt{MIN\_C} (blue circles) or \texttt{CBC} (orange squares). $R/L_{T,i}=7.5$ (solid line) and $R/L_{T,i}=5.0$ (dash-dot line).
  Markers are subsampled for clarity.}
  \label{fig:Fig_2}
\end{figure}

Figure~\ref{fig:Fig_2} presents the time traces of the normalized zonal free energy $\Xi$ for a highly turbulent state with $R/L_{T,i} = 7.5$ (solid lines)
and for a weakly turbulent state with $R/L_{T,i} = 5.0$ (dashed dot lines),  where zonal flows dominate.
\texttt{MIN\_C} simulations are represented by blue circular markers, while \texttt{CBC} simulations are shown with orange square markers.

For the high gradient case $R/L_{T,i}=7.5$, saturation occurs rapidly in less than $200\,R/C_s$,  and the saturation level is similar for both models with $\Xi \sim 10\%$.
This indicates that in strongly turbulent regimes, the choice  of the type of dissipation (i.e minimal zonal flow damping or not) have little influence on the overall free energy balance.

\begin{figure}[H]
  \includegraphics[width=\linewidth]{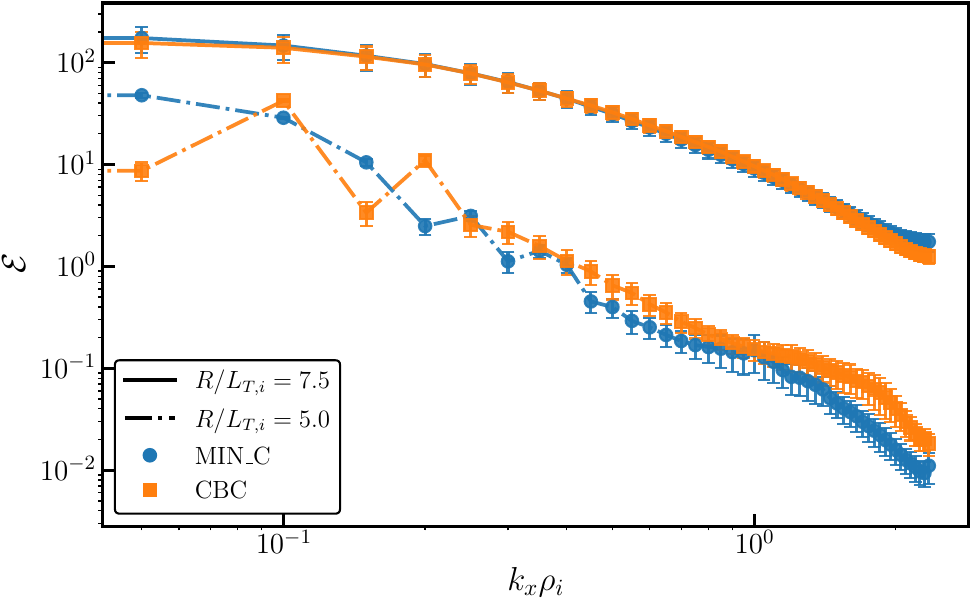}
  \caption{Free energy wave number $k_x$ spectra for \texttt{MIN\_C} (blue circles) and \texttt{CBC} (orange squares). $R/L_{T,i}=7.5$ (solid line) and $R/L_{T,i}=5.0$ (dash-dot line).}
  \label{fig:Fig_3}
\end{figure}

In contrast, for the weakly turbulent regime $R/L_{T,i}=5.0$, the simulations must run for at least $5000\,R/C_s$, corresponding to $\sim 1000\,\gamma^{-1}_{\max}$,  to reach saturated state.
In the CBC case, the zonal fraction saturates around $\Xi \sim 75\%$, while in the minimum zonal flow damping case, the level is higher, reaching $\Xi \sim 90\%$.
This discrepancy in saturation levels, observed only in the low temperature gradient regime, highlights the enhanced ability of the minimized 
set-up to promote the development of strong zonal flows.

Looking at these two distinct behaviors, we find that the simulations should be run for at least $1000\,\gamma^{-1}_{\max}$ to be sure that they reach saturation.
In the following, we choose $N=2000$ to ensure well-converged saturated states during our scans.

The spectra of free energy have also been analyzed and are presented in Figure~\ref{fig:Fig_3}.
Each spectrum is obtained by averaging the free energy over the saturated phase of the simulations, corresponding here to the interval $1000-2000\gamma^{-1}(R/L_{T,i})$. Error bars indicate standard deviation.

Similar to the time traces, the main differences between the spectra of the two choices of damping are observed in the zonal regime. 
For both \texttt{MIN\_C} and \texttt{CBC}, the overall level of the dominant mode is comparable, but the energy distribution across modes is different. 
In particular, \texttt{MIN\_C} exhibits more modes at higher energy levels, which correlates with larger levels of the order parameter $\Xi$ that is observed.
One drawback of minimizing the ZF dissipation is that energy tends to accumulate in small-scale structures, resulting in a slight increase of free energy at high $k_x$.

\subsection{Identification of Zonal, Intermediate, and Turbulent Regimes}

Having characterized the two extreme values of the temperature gradient, we turn to investigate the transition by considering a detailed scan in the interval $R/L_{T,i} \in \left [ 5.0, 7.5 \right ]$. 

\begin{figure}[H]
  \includegraphics[width=\linewidth]{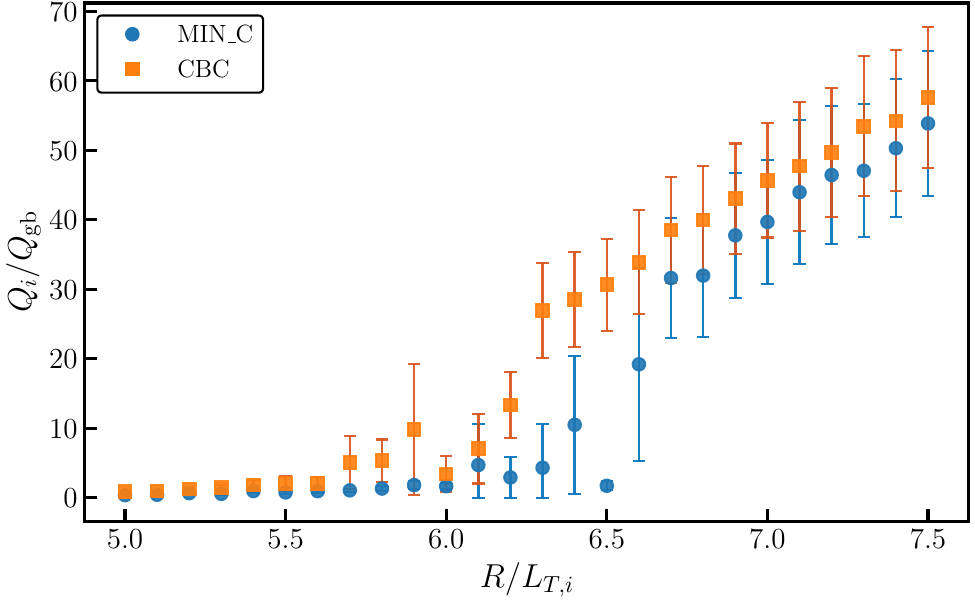}
 \caption{Averaged heat fluxes as a function of temperature gradient for \texttt{MIN\_C} (blue circles) or \texttt{CBC} (orange squares) parameters. Averages are taken in the interval $t \in \left [ 1000 ; 2000 \right ] \gamma^{-1}$, error bars are given by the associated standard deviations.}
  \label{fig:Fig_4}
\end{figure}

Figure~\ref{fig:Fig_4} presents the mean heat flux as a function of the temperature gradient $R/L_{T,i}$.
Each data point represents an independent simulation, where the heat flux has been averaged over the steady-state phase corresponding to the interval  $t \in \left [ 1000 ; 2000 \right ] \gamma^{-1}$,
where the error bars correspond to standard deviation.

\begin{figure*}[t]  
  \centering
  \includegraphics[width=0.96\linewidth]{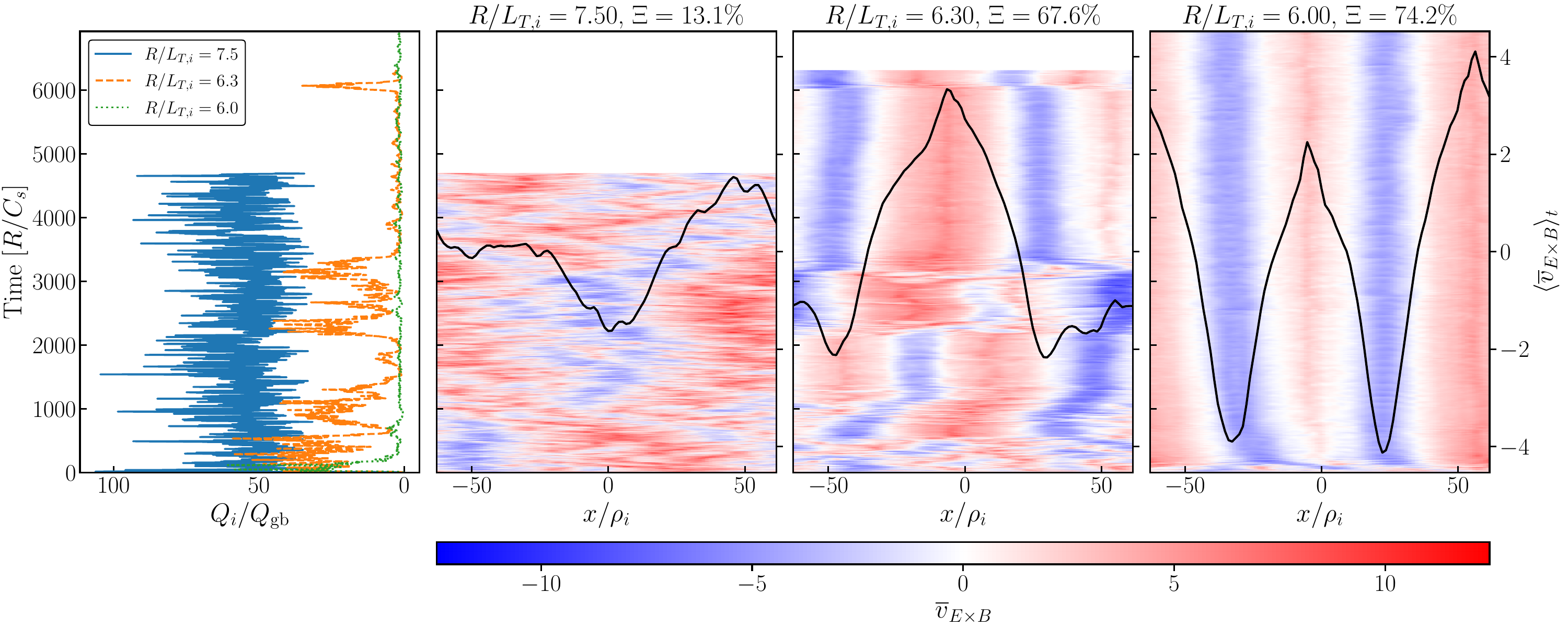}
  \caption{
  Left panel: ion electrostatic heat flux as a function of time for three temperature gradients, corresponding to
  turbulent  (solid blue line, $R/L_{T,i}=7.5$), intermediate (dashed orange line, $R/L_{T,i}=6.3$) and zonal regimes (dotted green line, $R/L_{T,i}=5.0$).
  Right panels: zonal $\overline{v}_{E\times B}$ velocity maps as a function of radial coordinate (horizontal $x/\rho_i$ axis) and time (vertical axis). From left to right: $R/L_{T,i}=7.5$ (turbulent regime), $R/L_{T,i}=6.3$ (intermediate regime), and $R/L_{T,i}=6.0$ (zonal regime). The black line corresponds to the time average zonal velocity $\langle \bar{v}_{E\times B} \rangle _t$}
  \label{fig:Fig_5}
\end{figure*}

As could be expected since the \texttt{MIN\_C} case corresponds to minimal zonal dissipation, the heat flux exhibits a shift toward higher values of $R/L_{T,i}$, as compared to the baseline case \texttt{CBC}.
This is consistent with previous observations, which show that the Dimits thresholds move to higher temperature gradients as the collisionality is reduced  \citep{Weikl2017}.
The observed levels of heat flux are similar for both the low and high gradient regimes, but presents a clear discrepancy in the transition region $R/L_{T,i}\in [6.1,6.6]$.
Focusing on the \texttt{MIN\_C} case, the heat flux remains stable in the zonal flow  regime and is nearly zero until the transition begins around $R/L_{T,i}=6.1$, where larger error bars are measured.
It then grows slowly with increasing $R/L_{T,i}$ up to about $R/L_{T,i}=6.6$, after which a sudden jump is observed, and the system enters the turbulent regime.
A similar trend is observed for \texttt{CBC}, with the transition occurring earlier, in the range $R/L_{T,i}=[5.7,6.2]$, which is also smoother and less abrupt.
Note that for $R/L_{T,i}=6.5$, a drop to a quasi-null heat flux state is observed, highlighting the fact that, close to marginality, the flux-gradient relation can be multi-valued.

Since it provides a more pronounced zonal flow dominated state, we consider the minimum zonal damping set-up \texttt{MIN\_C} in the remainder of this paper.

From the previous observations, three distinct regimes can be identified. To provide a clearer view of the three regimes, Figure~\ref{fig:Fig_5} shows independent simulations for $R/L_{T,i}=7.5$ (turbulent),
$6.3$ (intermediate), and $6.0$ (zonal), where the left panel presents the temporal evolution of the heat flux,
while the three right panels display the corresponding zonal velocity radial profile $\overline{v}_{E \times B}$ over time.
The mean radial velocity profile $\langle \bar{v}_{E\times B} \rangle _t$  is also given by the black curve, whose value is given by the right $y$-axis.

The three different regimes can be described as follows:

\begin{itemize}
  \item \emph{Dimits / Zonal regime} ($R/L_{T,i} \leq 6.0$):  
  The heat flux is nearly zero after the initial linear overshoot. 
  In real space, the zonal velocity profile $\overline{v}_{E \times B}$ exhibits persistent, temporally stable bands of nearly constant radial widths, alternating between positive and negative values. 
  These structures present strong, radially sheared, zonal flows related to the \emph{staircase} \citep{DifPradalier2015, Lippert2023} structures in flux-driven simulations,
  which suppress turbulence once a sufficiently strong and stable sheared flow has developed across the radial domain.

  \item \emph{Turbulent regime} ($R/L_{T,i} \geq 6.7$):  
  The heat flux fluctuates strongly around a large mean value.
  The corresponding zonal velocity profile is also highly fluctuating both spatially and temporally, suggesting that we are considering the zonal component of a properly turbulent velocity field.

  \item \emph{Intermediate regime} ($6.1 \leq R/L_{T,i} \leq 6.6$):  
  The heat flux exhibits intermittent behavior alternating between a finite turbulent phase and a quasi-null zonal-dominated phase. 
  For instance, at $R/L_{T,i} = 6.3$,
  the system alternates between periods of strong turbulence ($0 < t < 1300,R/C_s$, $2100 < t < 3400,R/C_s$ and $6000 < t < 6200,R/C_s$) and intervals of suppressed turbulence dominated by strong zonal flows.
  The heat flux and the zonal flow velocity profile evolve coherently and simultaneously.
  A drop of the heat flux to quasi-null values coincides with the formation of banded structures in the zonal velocity profile, indicating a zonal-dominated state.
  Conversely, bursts of finite heat flux are associated with the destabilisation/destruction of these bands and the re-emergence of small regions with fast variations of velocity.
  These oscillations occur at low frequencies, on the order of a few hundred to several thousand $R/C_s$'s and prevent the heat flux from converging to a well-defined steady state value.  
\end{itemize}

Overall, it becomes apparent that the system can be found in two distinct dynamical states, which may suggest the presence of two distinct "thermodynamical phases",
with the possibility of a transition between them, even though the finite system with a particular form of dissipation that we consider, in a limited resolution,
will never properly represent a true thermodynamical phase transition.
Nonetheless, we can use the zonal free energy fraction $\Xi$, as an order parameter quantifying the zonal contribution to the total free energy of the system, in order to characterize the underlying transition.
Using this parameter, the different regimes can be distinguished roughly as:

\begin{equation}
\begin{cases}
  \Xi > 0.70 &\Rightarrow \emph{Zonal regime}, \\[6pt]
  0.30 \leq \Xi \leq 0.70 &\Rightarrow \emph{Intermediate regime}, \\[6pt]
  \Xi < 0.30 &\Rightarrow \emph{Turbulent regime}.
\end{cases}
\end{equation}

\begin{figure}[H]
  \centering
  \includegraphics[width=\linewidth]{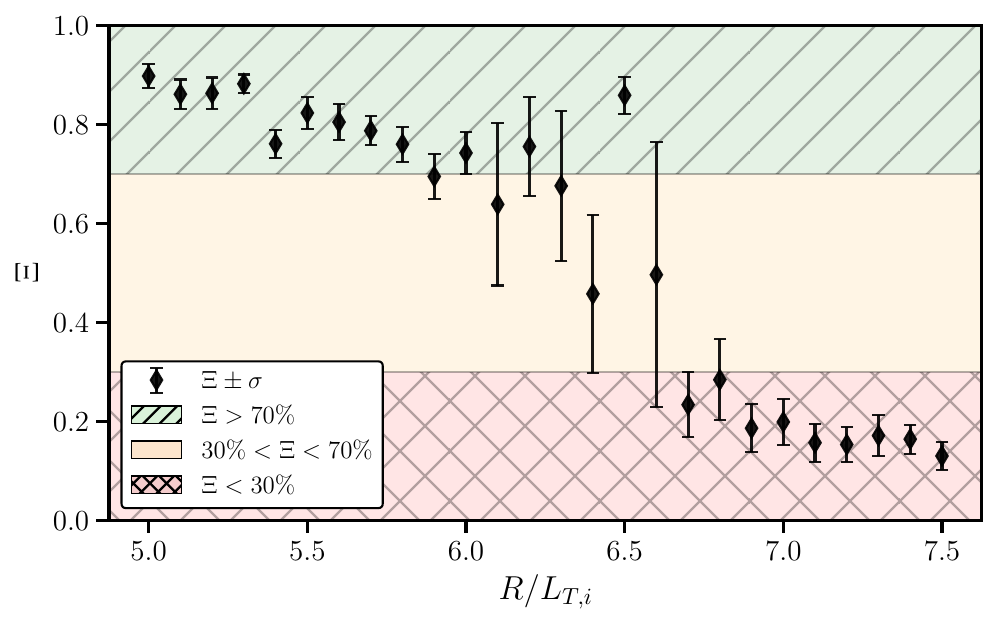}
  \caption{Zonal free energy fraction $\Xi$ as a function of temperature gradient. Zonal flow dominated ($\Xi > 70\%$, green with diagonal hatching), intermediate regime ($30\% < \Xi < 70\%$, orange with dotted hatching), and turbulent regime ($\Xi < 30\%$, red with cross hatching) regimes are highlighted.}
  \label{fig:Fig_6}
\end{figure}

These regimes are shown in Figure~\ref{fig:Fig_6}, which presents the averaged zonal free energy fraction $\Xi$ over the saturated phase $1000$-$2000\gamma^{-1}$,
with error bars representing the corresponding standard deviation. 
To visually distinguish the different regimes, two hatch patterns are used in combination with colors: the zonal flow dominated regime ($\Xi > 70\%$, green with diagonal hatching),
the intermediate regime ($30\% < \Xi < 70\%$, orange), and the turbulent regime ($\Xi < 30\%$, red with cross hatching).
Consistently with the results obtained using the heat flux, the ranges for each regime are recovered: $R/L_{T,i}=[5.0,6.0]$ for the zonal regime $R/L_{T,i}=[6.1,6.6]$, for the intermediate regime,
and $R/L_{T,i}=[6.7,7.5]$ for the turbulent regime.

Note that the intermittent behavior of the intermediate regime becomes more evident using the order parameter $\Xi$.
In this regime, $\Xi$ exhibits a very large standard deviation, reflecting the back and forth between zonal-dominated and turbulent phases,
with the exception of $R/L_{T,i}=6.5$ (see Fig. \ref{fig:Fig_10} bottom left panel black curve), which remains in a zonal flow dominated state.
This strong variability suggests that the system fluctuates between two "fixed points" from a dynamical systems point of view (Left panel \ref{fig:Fig_5} heat flux time trace for $R/L_{T,i}=6.3$),
which could imply bistability. These oscillations are indicators of a strong interplay between turbulence and ZFs, often described using a predator-prey analogy.

\section{Chained Simulations and Independent Simulations \label{sec:ramps}}

Since the independent simulations show that in the intermediate regime, the system exhibits oscillatory behavior between two states, suggestive of bistability,
we decide to explore this transition region in more detail to examine the existence and the role of memory effects inherited from previous states.

\begin{figure}[H]
  \includegraphics[width=\linewidth]{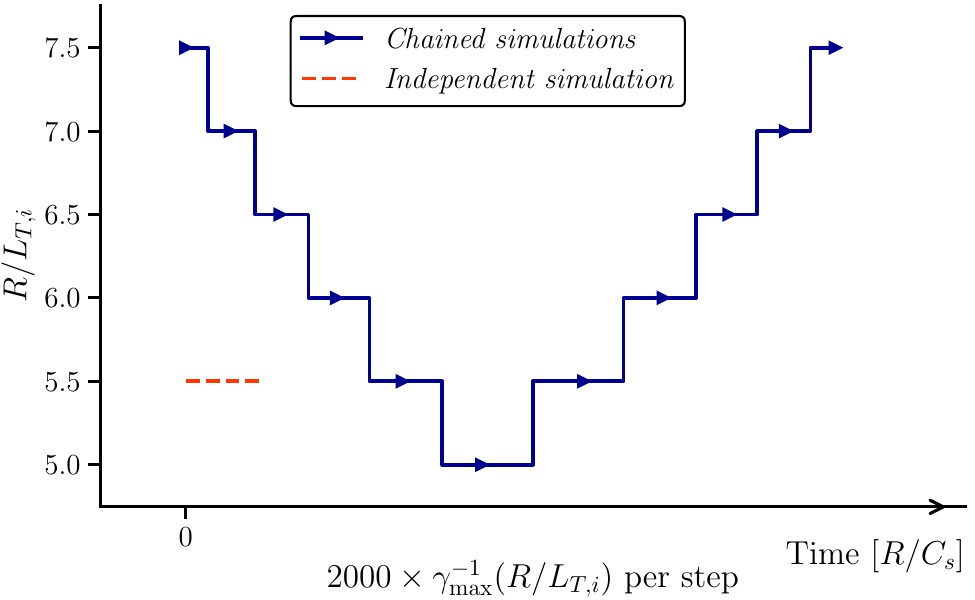}
  \caption{Schematic representation of chained simulations with a step $\Delta R/L_{T,i}=0.5$ (blue, solid line). $R/L_{T,i}$ is varied stepwise after a fixed simulation time determined by the number $N \times \gamma^{-1}_{\max}$, where the growth rate depends onto the temperature gradient $\gamma_{max} = \gamma_{max} (R/L_{T,i})$.}
  \label{fig:Fig_7}
\end{figure}

To this end, we perform chained simulations, as described schematically in Figure~\ref{fig:Fig_7}, and compare them to previous independent simulations.

Notice that in the independent simulations, each simulation is initialized with zero initial conditions plus small random perturbations at a fixed temperature gradient.
The system then evolves freely through a linear growth phase before reaching nonlinear saturation.
In contrast, the chained simulations are performed using a \emph{restart} procedure: each simulation of the chain is initialized from the final state of the previous one, while the temperature gradient 
$R/L_{T,i}$ is modified from one step to the next.
Specifically, at the end of the $j$th step of the ramp with temperature gradient $R/L_{T,i}(j)$,  the $(j+1)$th step starts from the final state of the previous step, but with the updated gradient $R/L_{T,i}(j+1)$.
The first simulation of the chain is initialized in the same way as an independent one, with small random perturbations.
Such an approach preserves the memory of previously developed zonal flows, allowing the investigation of how the persistence of such structures influence the evolution of the system.  

\subsection{Total, Zonal, and Turbulent free energy time traces in Chained vs. Independent Simulations}

\begin{figure}[H]
  \includegraphics[width=\linewidth]{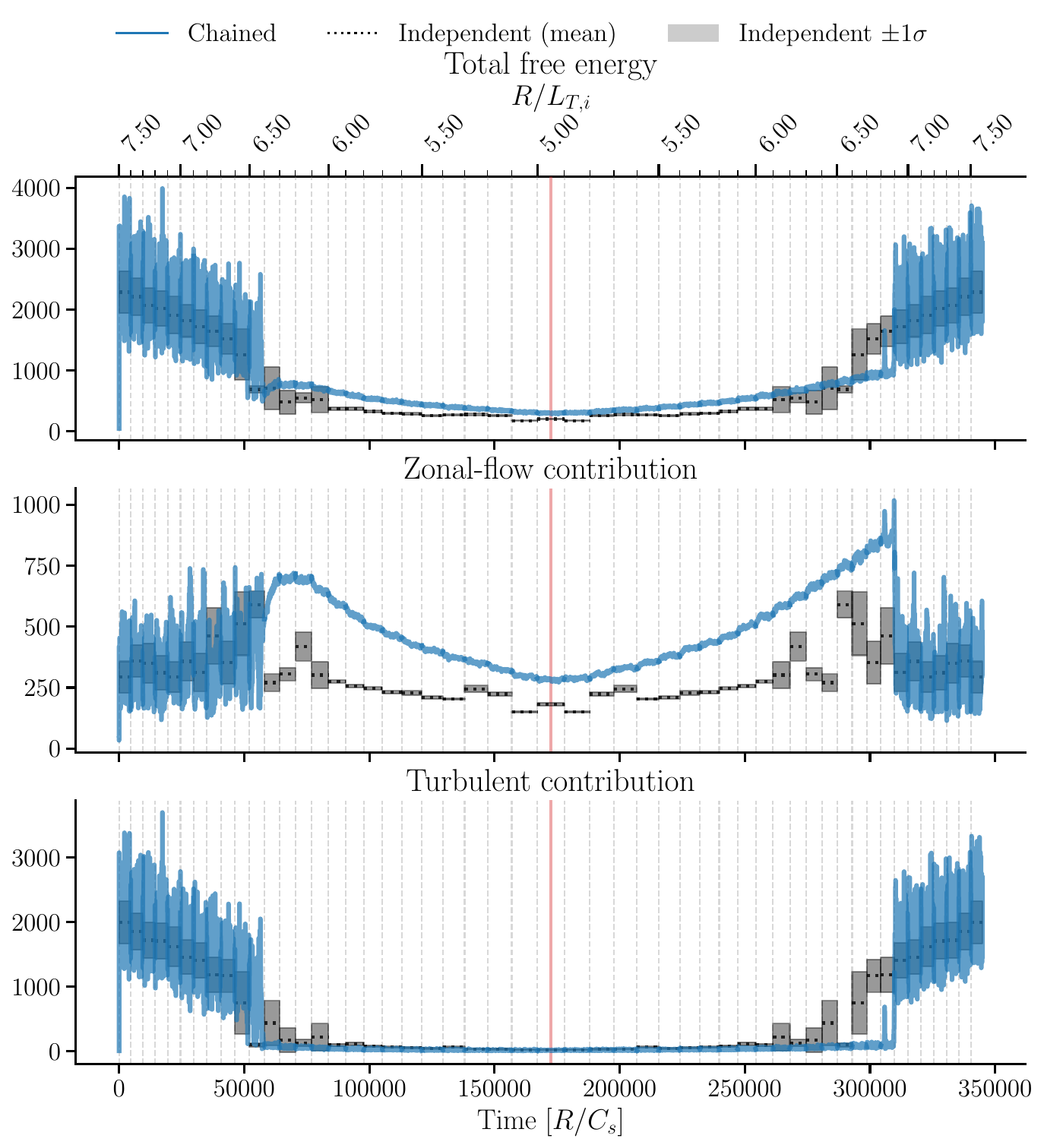}
  \caption{Total, zonal, and turbulent free energies, as functions of time (bottom horizontal axis), and temperature gradient (top horizontal axis), as obtained from chained simulations (blue, solid, line), and compared to independent simulations (black dotted for mean value for a given gradient $R / L_{T_i}$, errorbars from standard deviation).}
  \label{fig:Fig_8}
\end{figure}

In the comparison between chained to independent simulations, we first focus on the effect of a gradual decrease of the temperature gradient, corresponding to the left half of Figure~\ref{fig:Fig_8}. 

Figure~\ref{fig:Fig_8} shows the time evolution of free energy for a chained (solid blue line) and independent simulations (dotted black line) at fixed temperature gradients. 
For the independent simulations, the mean values over the steady state are shown with the shaded area representing standard deviation.
The independent simulations are symmetric by definition, which is nonetheless displayed for both the ramp up and the ramp down to compare against the chained simulation.
A secondary $x$-axis at the top of the first panel indicates the evolution of the temperature gradient  $R/L_{T,i}$ over time.
The first panel displays the total free energy, while the second and third panels show the zonal Eq.\eqref{eq: time zonal free energy contribution}
and turbulent Eq.\eqref{eq: time turbulent free energy contribution} contributions, respectively.

The results from chained and independent simulations are nearly identical for the turbulent regime and part of the intermediate regime close to it (i.e. $R/L_{T,i} \geq 6.5$).
In contrast, the zonal flow contribution is higher in the chained simulations for the rest of the intermediate regime and in the Dimits shift, as $R/L_{T,i}$ decreases (i.e. for $R/L_{T,i} \leq 6.4$). 
This pronounced discrepancy between independent scans vs. ramp down to lower temperature gradients can be attributed to the persistence of zonal flows that are generated during previous steps
(i.e. at higher gradients) in steady state. When we decrease the gradient further, these existing established zonal flows are probably more efficient to extract energy from turbulence.

Note that, zonal flows increase and their intensity reaches its maximum around $R/L_{T,i} = 6.4$-$6.2$, corresponding to the intermediate regime. 
At this point, they are sufficiently strong to maximize their interaction with turbulence,  feeding on the available free energy. This interaction is also reflected in the turbulent contribution: 
the net turbulence level is higher in the independent simulations, where less zonal flows develop and therefore cannot regulate turbulence as effectively as in the chained case.

As the temperature gradient continues to decrease, the zonal flow contribution weakens due to reduced turbulent drive.
Such a decrease may be partly related to dissipation. 
However, once the system transits to a zonal dominated state with nearly vanishing turbulence,
both the dissipation and the nonlinear transfer terms in the free energy balance decrease significantly. 
Therefore, the reduction of the zonal flow free energy cannot be unambiguously attributed to dissipation alone,
and may instead result from a combination of reduced nonlinear transfer and dissipative effects.
Beyond the peak of zonal flow intensity (i.e. around $R/L_{T,i} \leq 6.1$), 
the turbulent contributions in the independent and chained simulations recover comparable levels and eventually coincide,
while the zonal flow component in the chained simulations remains slightly elevated, 
indicating a memory of the previous step, and the possibly multi-valued nature of the zonal flow dominated state.

This notion of residual memory is particularly interesting, as it appears to push the system toward a strongly zonal flow dominated state.
Such a behavior may reflect the multivalued nature of the system deep within the Dimits regime, 
where turbulence eventually decays and the zonal flow amplitude is not uniquely defined,
but can settle at different levels depending on the transient evolution of the system.

Then, we consider the right half of Figure~\ref{fig:Fig_8}, corresponding to a ramp up of the temperature gradient starting from a low-turbulence state dominated by strong zonal flows.

As the temperature gradient increases, the free energy available to drive turbulence also increases.
However, since a significant fraction of this energy is stored in zonal flows, the turbulence cannot develop initially, and this additional free energy is also predominantly transferred to zonal flows,
resulting in their further amplification within the range $R/L_{T,i} = [5.0, 6.8]$.

Furthermore, this growth phase extends over a broader interval than the ramp down, where the maximum zonal free energy is reached around $R/L_{T,i} = [6.4, 6.2]$.
When increasing toward $R/L_{T,i} = [6.2, 6.4]$ the zonal flow amplitude reaches a magnitude comparable to the ramp down, but instead of peaking, this level keeps increasing until $R/L_{T,i} = 6.8$,
with the turbulence still mostly suppressed.
Two overshoots in zonal flow and turbulence levels are observable at these gradient values, just before the collapse of the zonal flow. Then, abruptly, the free energy injection to turbulence becomes high enough,
and the turbulence starts to grow again, leading to a sudden collapse in the zonal flow free energy.
After the transition back to turbulence regime, turbulence recovers its previous level and matches, once again, the independent simulations. 

In short, a clear asymmetry is observed between the decreasing and increasing gradient phases,
suggesting a path dependence of the system and a possible hysteresis in the response of zonal free energy to temperature gradient.

\subsection{Evolution of the order parameter $\Xi$ in chained simulations.}

Since we observe some kind of hysteresis in the response of zonal free energy, to slow variation of the temperature gradient,
we consider using the zonal free-energy fraction $\Xi$ as an order parameter to better characterize the state of the system for studying hysteresis.

\begin{figure}[H]
  \includegraphics[width=\linewidth]{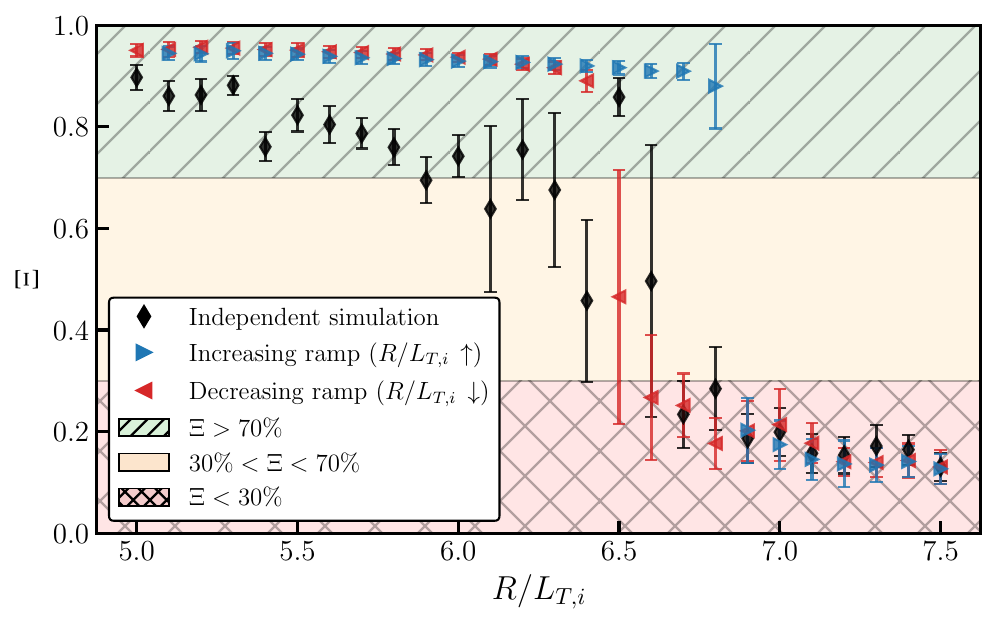}
  \caption{Zonal free energy fraction as a function of temperature gradient, as obtained from chained simulations (decreasing or increasing temperature gradients symbolized respectively by red left  or blue right pointing triangles), and compared to independent simulations (black diamonds).}
  \label{fig:Fig_9}
\end{figure}

Figure~\ref{fig:Fig_9} shows $\Xi$ as a function of the temperature gradient $R/L_{T,i}$.
Here, $\Xi$ is averaged over the full duration of each temperature gradient step for the chained simulations, and over the steady state for the independent simulations,
and the error bars represent standard deviation over the interval.
The ramps are shown using triangle markers, with the direction of the symbol indicating the direction of the temperature gradient change (i.e. pointing right means gradient is being increased),
with independent simulations shown using diamond black markers. Note that the data here are the same as in Figure~\ref{fig:Fig_6}, but here the chain starts on the lower right on the red curve,
ramp down to top left and then ramp back up, following the blue curve. 

As discussed previously in  Figure~\ref{fig:Fig_6} and now denoted by black diamond shaped markers in Figure~\ref{fig:Fig_9},
the independent simulations exhibit strong fluctuations of $\Xi$ in the range $R/L_{T,i} \in [6.1, 6.6]$, reflected by large standard deviations.
This behavior indicates that the system alternates between two marginally stable states, corresponding to turbulent and zonal-dominated configurations.

In contrast, decreasing the temperature gradient in chained simulations, represented by the red left-pointing triangle markers '{\color{red} \rotatebox[origin=c]{90}{$\blacktriangle$}}',
the system remains near the zonal flow dominated fixed point, with a significantly reduced standard deviation. This suggests that the system retains some kind of memory, and an underlying bistability.
One alternative, but possibly related explanation is that, the time spent in the turbulent phase allows the system to deplete small scale "enstrophy" (here it would be the square of plasma vorticity),
so once the turbulence is suppressed, the zonal flow appears as a robust minimum enstrophy state.

Notice that, it is only near the transition point at $R/L_{T,i} = 6.5$ that we can observe strong variations, leading to larger error bars in chained simulations.
This is consistent with the onset of the transition, where the system undergoes significant dynamical changes.
After the transition, the chained simulations reach a strongly zonal-dominated state, with $\Xi = 0.90$-$0.95$,
well in excess of the values obtained in independent simulations ($\Xi = 0.70$-$0.90$ for $R/L_{T,i} \in [5.0, 6.0]$).

For the chained simulations, where the temperature gradient is ramped back up, 
depicted by right-pointing blue triangle markers '{\color{blue} \rotatebox[origin=c]{270}{$\blacktriangle$} }',
$\Xi$ remains at a higher value over a wider range of temperature gradients ($R/L_{T,i}\in [5.0,6.8])$ compared to the ramp down phase ($R/L_{T,i}\in [6.5,5.0]$),
highlighting the persistence of the zonal dominated state once it is established.
Note the analogy to crystal melting and the concept of latent heat, 
where the back transition to turbulence (disordered, liquid like state) occurs only when sufficient energy is supplied to the system 
to break down the zonal structures, which can be viewed as plasma analogues of the solid phase.

\begin{figure*}[t]  
  \centering
  \includegraphics[width=0.80\textwidth]{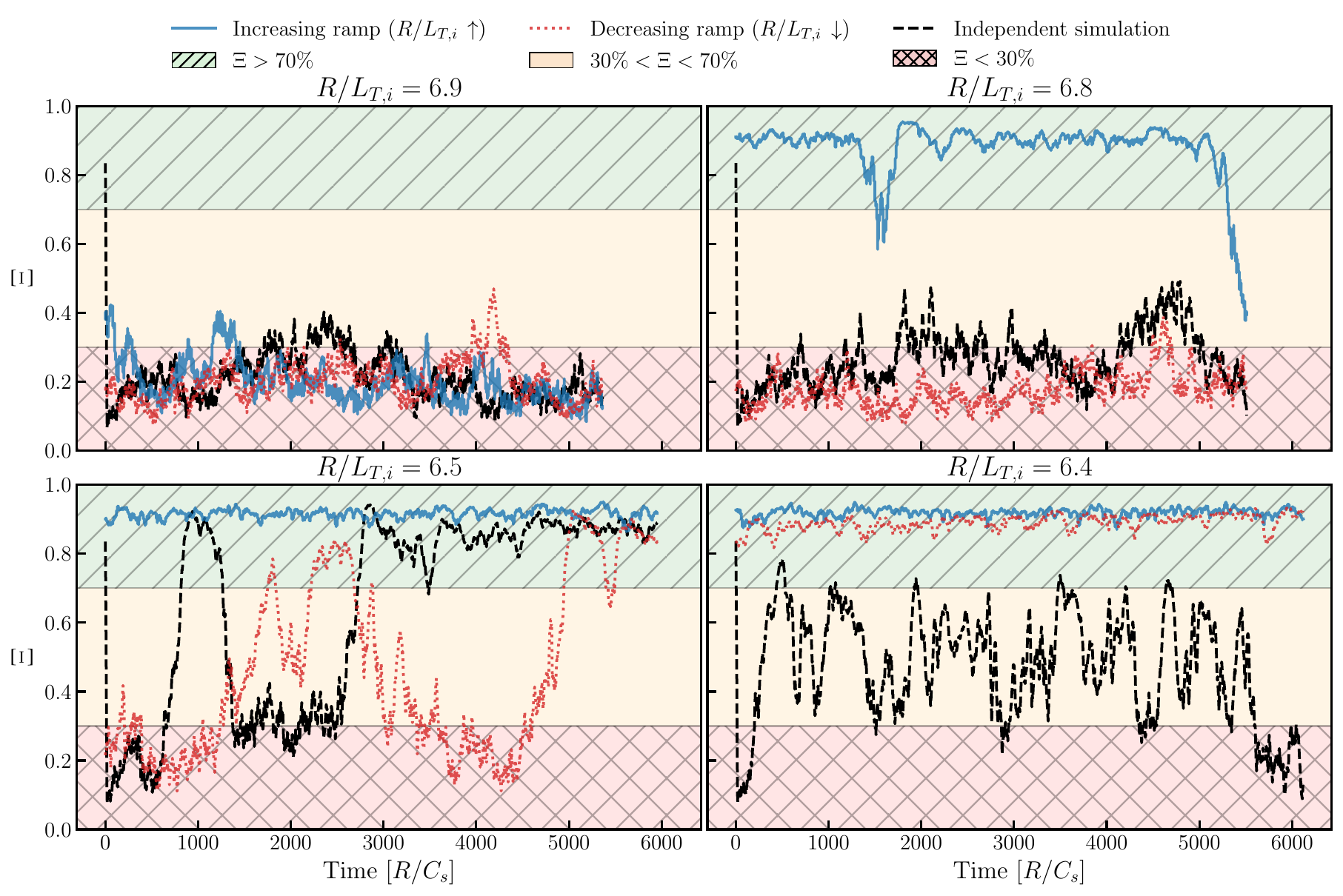}
  \caption{
  Zonal free energy fraction as a function of time, as obtained from independent (black dashed line), and from chained simulations respectively decreasing (red dotted) and increasing (blue line) the temperature gradient, for four representative value of the temperature gradient ($R/L_{T_i} = 6.9$, $6.8$, $6.5$ and $6.4$ from top left to bottom right panels).
  }
  \label{fig:Fig_10}
\end{figure*}

To gain further insight into the transition, Figure~\ref{fig:Fig_10} shows the time evolution of the order parameter $\Xi$ for both the chained and independent simulations at four representative values of $R/L_{T,i}$.
The decreasing phase of the chained simulation is shown in dotted red lines and the increasing one is shown in blue,
while the independent simulation are represented by dashed black lines.
Here, the top left panel showing $R/L_{T,i}=6.9$ and the bottom right panel showing $R/L_{T,i}=6.4$ correspond to the turbulent and zonal regimes, respectively,
while bottom left panel showing $R/L_{T,i}=6.5$ illustrates the transition from turbulence to zonal dominated state, and the top right panel showing $R/L_{T,i}=6.8$ shows the transition back to turbulence.

The transition from turbulence to a zonal flow dominated state ($R/L_{T,i}=6.5 \rightarrow 6.4$), indicated by the red dotted line, occurs gradually.
The system intermittently switches between zonal and turbulent states and exhibits low-frequency oscillations before settling.
Reducing the temperature gradient to $R/L_{T,i}=6.4$ stabilizes the zonal state by weakening the linear drive.
Note that for the $R/L_{T.i}=6.4$ the independent simulation shows a clear and pronounced intermittent behavior.

The reverse transition ($R/L_{T,i}=6.8 \rightarrow 6.9$), indicated by the solid blue curves, differs in that it is triggered by short-lived turbulent bursts.
Two such events are observed, with the second being sufficiently strong to drive the system back to the turbulent regime, where the zonal free energy fraction remains low.

Overall, the two transitions occur on distinct timescales: the formation of a zonal-dominated state is gradual, whereas the transition back to turbulence can occur rapidly once a sufficiently strong burst develops.

\section{Wave number Spectra\label{sec:spectra}}

We now consider the implications of the transition from turbulent to zonal dominated states on the wave-number spectra to see if it has any scale dependence.
Here, we consider the $k_x$ spectra, because the zonal flows are expected to scatter free energy in $k_x$, and it allows us to investigate the zonal flow spectrum as well.  

At a given time $t$, $k_x$-spectrum of free energy can be written as: 
\begin{equation}
  \label{eq:full_kx-spectra}
  \mathcal{E}(k_x,t)=2 \times \sum_{k_y > 0} \mathcal{E}(k_x,k_y,t)+\mathcal{E}(k_x,k_y=0,t) \, , 
\end{equation} where the first term of Eq.\eqref{eq:full_kx-spectra}
corresponds to the turbulent contribution given by Eq.\eqref{eq: time turbulent free energy contribution} and the second term gives the zonal contribution Eq.\eqref{eq: zonal free energy contribution}

In order to represent average spectrum, $\mathcal{E}(k_x)=\langle \mathcal{E}(k_x,t)\rangle_T$,
independent simulation are averaged over the nonlinear saturation phase to skip the linear transient while the chained simulations are averaged over the entire time interval associated to each $R/L_{T,i}$ step.

\subsection{Spectra associated to turbulent and zonal regimes}

\begin{figure}[H]
  \centering
  \includegraphics[width=\linewidth]{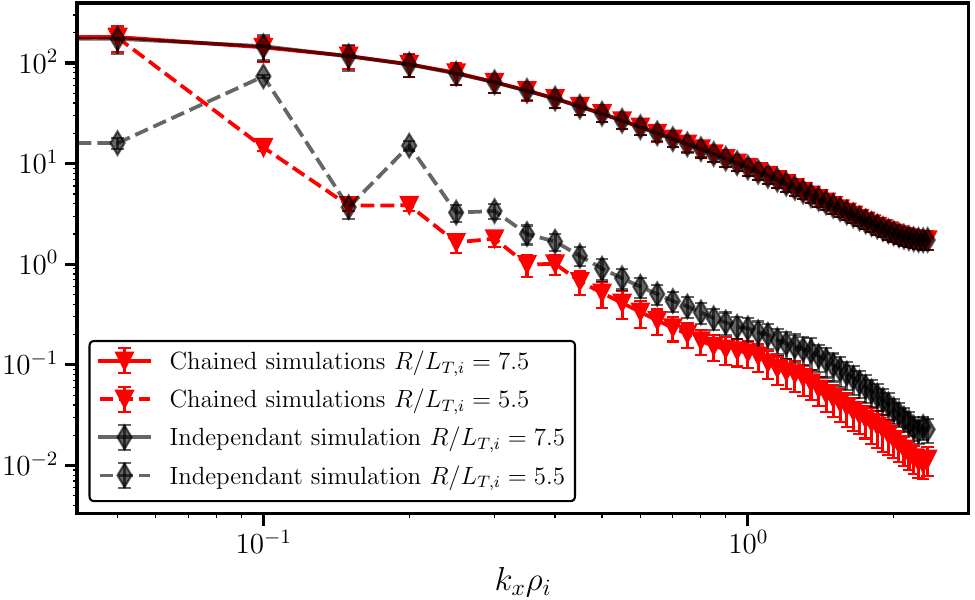}
  \caption{
  Free energy $k_x$ spectra from chained (red downward triangles) and independent simulations (black diamonds), corresponding to the turbulent regime $R/L_{T,i}=7.5$ (solid lines), and to the zonal dominated regime $R/L_{T,i}=5.5$ (dashed lines).}
  \label{fig:Fig_11}
\end{figure}

\begin{figure*}[t]
  \centering
  \includegraphics[width=0.75\linewidth]{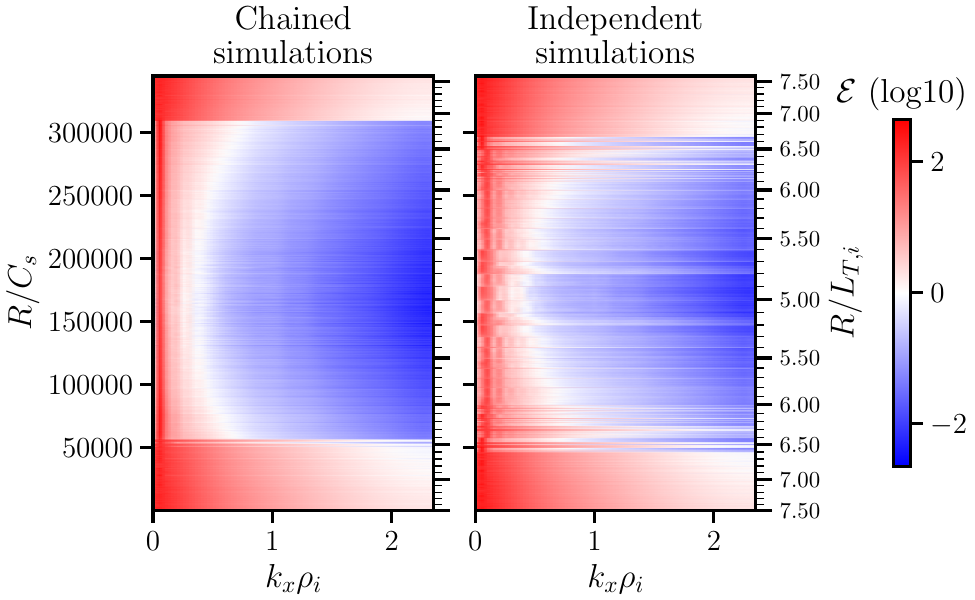}
  \caption{
  Time evolution of the free energy $k_x$ spectra, smoothed with a linear uniform filter, as a functions of time (left $y$-axis) and temperature gradient (right $y$-axis). Results from chained (left panel) and independent (right panel) simulations are shown. Independent simulations are symmetrized around $R/L_{T,i}=5.0$ to facilitate comparison with the first and second halves of the ramp in the chained case.}
  \label{fig:Fig_12}
\end{figure*}

Figure~\ref{fig:Fig_11} shows the $k_x$ spectra of free energy for $R/L_{T,i}=7.5$ in the turbulent regime (solid lines) and $R/L_{T,i}=5.5$ in the  zonal regime (dashed lines).
Chained simulations are represented by red downward triangles indicating that the temperature gradient has been decreased from the previous state. Independent simulations are represented by black diamonds.
Since the increasing and decreasing paths of the chained simulations produce similar spectra deep in both the zonal and turbulent regimes, only the decreasing phase is presented here.

The first difference between the two regimes is in the overall magnitude: 
the wave number spectra in the turbulent case are approximately one order of magnitude larger than the zonal flow dominated case.
This behavior is consistent with the time evolution of the total free energy shown in Figure~\ref{fig:Fig_8},
where a drop from $1000$-$4000$~A.U. to about $500$~A.U. is observed while transitioning from the turbulent to the zonal state.

The spectral shapes also differ significantly between the two regimes.
The turbulent one shows roughly two slopes that join around $k_x\rho_i=0.5$: a slow decaying slope at large scales a steeper slope at small scales.
In the zonal dominated case, the energy is peaked at large scales and decays significantly faster than the turbulent case. Note also that for the large scales,
the peak for the independent simulation is the second largest available mode (half the box size) whereas it is the largest available one, with $k_x\rho_i=0.05$, in the case of chained simulations.

This distinction is not specific to the value $R/L_{T,i}=5.5$ and can be observed for values of the temperature gradient associated with the zonal flow dominated regime,
and likely explains the differences observed in Figure~\ref{fig:Fig_9} between the zonal flow dominated regimes of chained vs. independent simulations.

\subsubsection{Dynamical behavior of the $k_x$ spectra.}

To understand the dynamics of the wave-number spectrum of free energy we use a moving-average representation in Figure~\ref{fig:Fig_12},
where time and the corresponding $R/L_{T,i}$ are shown on the left and right $y$-axes with the left and the right panels showing chained and independent simulations respectively.

Two distinct phases can be identified for the chained simulations represented on the left panel of Figure~\ref{fig:Fig_12},
where the temperature gradient is gradually decreased from $R/L_{T,i}=7.5$ down to $R/L_{T,i}=5.0$, then increased back up to $R/L_{T,i}=7.5$.
At high temperature gradients, corresponding to the intervals $R/L_{T,i}=7.5 \rightarrow 6.5$ at the bottom and $R/L_{T,i}=6.8 \rightarrow 7.5$, at the top, the system is in a turbulent regime, which is characterized by extended high-amplitude regions that decay smoothly toward larger $k_x\rho_i$ values (red to white in color online), consistent with the two-slope spectral structure identified in the previous section. In contrast, the central part, corresponding to the zonal flow dominated regime, exhibits a strong concentration of free energy at small $k_x$, with a dominant contribution at $k_x\rho_i = 0.05$. This indicates that most of the free energy is stored in large scale zonal flows.

Note that, the dynamics of the spectrum also exhibit an asymmetry with slightly different spectral widths for the same gradient depending on if it is in the ramp down or the ramp up.
We also find that the transition from turbulence to zonal flows occurs near $R/L_{T,i}=6.5$ while the transition from the zonal flows back to turbulence occurs around $R/L_{T,i}=6.8$.

\begin{figure*}[t]
  \centering
  \includegraphics[width=0.75\linewidth]{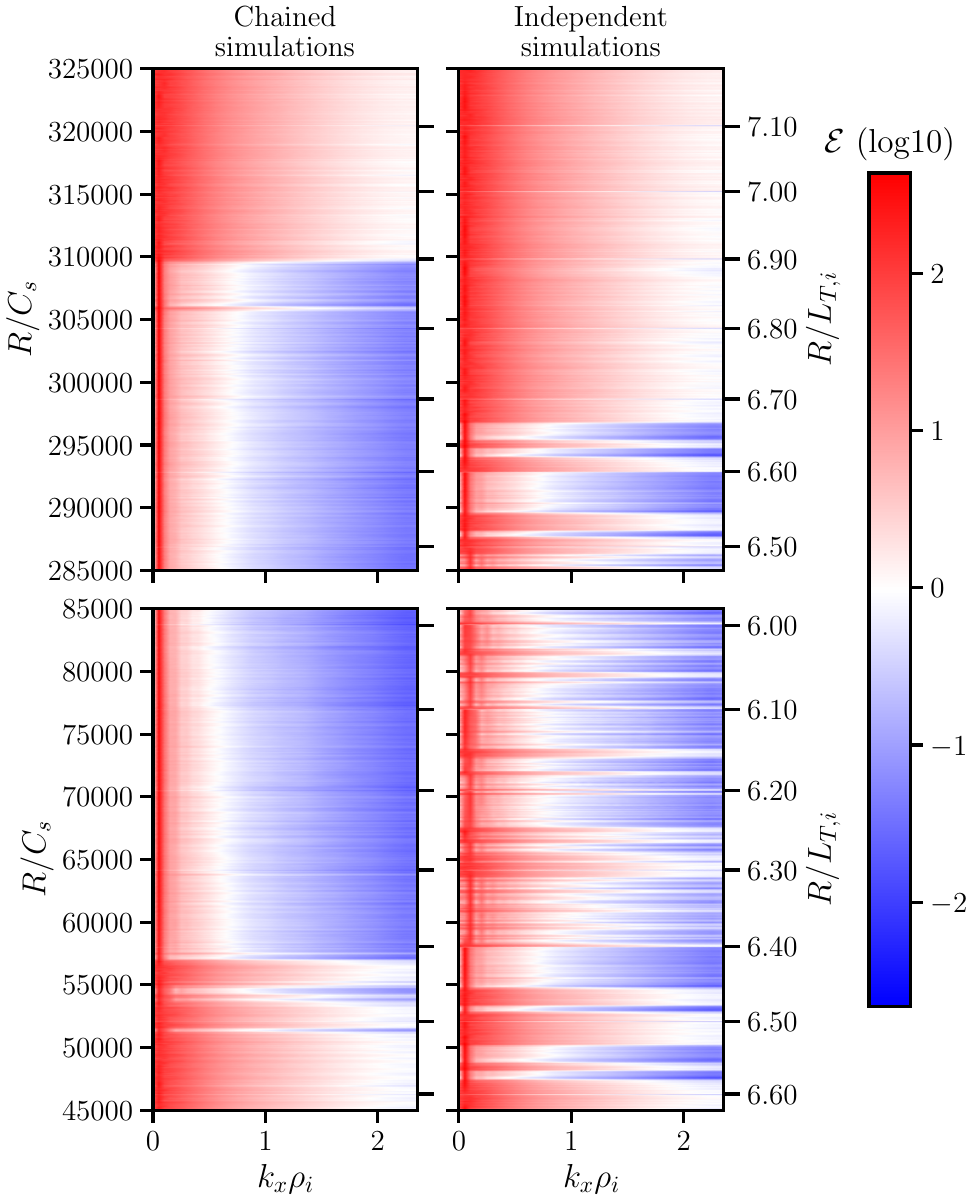}
  \caption{
  Zoomed-in view of Figure~\ref{fig:Fig_12} near the transition regions. Top row: transition from the zonal flow dominated state to the turbulent state, $t \in [285000,325000]$ ($R/L_{T,i} \in [6.5 \rightarrow 7.1]$). Bottom row: transition from the turbulent to the zonal flow dominated state, $t \in [45000,85000]$ ($R/L_{T,i} \in [6.6 \rightarrow 6.0]$). Left panels: chained simulations, right panels: independent simulations.}
  \label{fig:Fig_13}
\end{figure*}

In contrast, for the independent simulations shown on the right panel, the zonal regime ($R/L_{T,i}\in[5.0,6.0]$) does not seems to exhibit a uniquely dominant mode.
Depending on the value of the temperature gradient, the free energy may peak at $k_x\rho_i = 0.05$, at $k_x\rho_i = 0.10$, or be distributed between these two modes (e.g.\ $R/L_{T,i}=5.2$-$5.3$),
and in some cases even between $k_x\rho_i = 0.10$ and $k_x\rho_i = 0.20$ (for $R/L_{T,i}=5.5$-$6.0$).
In contrast, for the chained simulations the zonal flow is systematically characterized by a dominant contribution at $k_x\rho_i = 0.05$.
This highlights the robustness of the large-scale zonal flow structure once it has formed and is maintained over time.

Another notable feature of the independent simulations is the noisy temporal behavior of the spectra going back and forth between turbulent
and zonal dominated states when the temperature gradient lies in the intermediate regime $R/L_{T,i}\in[6.1,6.6]$.
Such intermittent switching is not observed in the chained simulations, highlighting, once again, the crucial role of memory effects.

\subsection{Details of the transition}

A closer view of the dynamics of the free energy $k_x$ spectra across the transition is provided in Figure~\ref{fig:Fig_13}, 
which gives a zoom around the transition regions observed into the time evolution shown Figure~\ref{fig:Fig_12}.

Here, the top row shows the transition from the zonal to turbulent state, for the interval $t\in[285000,325000]\, R/C_s$, corresponding to $R/L_{T,i}\in[6.5 \rightarrow 7.1]$,
while the bottom row shows the transition from the turbulent to zonal state,
over the interval $t\in[45000,85000]\, R/C_s$, corresponding to $R/L_{T,i}\in[6.6 \rightarrow 6.0]$.
In both cases, the left column shows the chained simulation, while the right column shows the independent simulations.

As the temperature gradient is decreased from $R/L_{T,i}=7.5$, the first transition from a turbulent to a zonal flow dominated state occurs around $R/L_{T,i}=6.6$,
which is shown at the left panel of the bottom row of Figure~\ref{fig:Fig_13}.
During this step, the spectrum remains mostly in a turbulent state until towards the end of the step (i.e. a bit above $t=50000~R/C_s$),
a brief incursion into the zonal flow dominated state can be noticed. After multiple attempts at $R/L_{T,i}=6.5$, the transition to the zonal flow dominated state is observed just before $R/L_{T,i}=6.4$.

Beyond these steps, the spectra remain stable as $R/L_{T,i}$ is varied from $6.3 \rightarrow 5.0 \rightarrow 6.7$ (see Figure~\ref{fig:Fig_12} for complete view).
This corroborates the conclusion once established, that the zonal flow dominated state is rather persistent in wave-number spectra as well.

In contrast, independent simulations (right panel) exhibit numerous switchover events within the same temperature gradient interval.
This suggests that the zonal flows do not build up sufficiently to suppress the turbulence as effectively in independent simulations as in the chained simulations,
where the zonal structures are inherited from previous steps.

The top row of Figure~\ref{fig:Fig_13} illustrates the transition from the zonal flow dominated state back to turbulence.  
The first perturbation occurs during the time interval corresponding to $R/L_{T,i}=6.8$, where the free energy, initially concentrated in the mode $k_x\rho_i=0.05$,
is temporarily redistributed to smaller scales, producing a broadband spectrum briefly, before returning to a state dominated by the $k_x\rho_i=0.05$ mode.
Then towards the end of this step, a bit before $t = 310000~R/C_s$, the system completely transits to a turbulent state, as indicated by the broadband distribution of free energy across $k_x$.

\subsection{Characterization of large scale spectral slope of the $k_x$ spectra}

Examining the dynamics of the $k_x$ spectra, the distinction between the two regimes is primarily in the distribution of energy at large scales and the spectral slope.
Hence one can instead focus on the spectral range $k_x\rho_i \in [0.05,0.5]$ corresponding to the first ten $k_x$ modes of the system to quantify the difference of behavior in these two regimes,
considering a power law scaling for these scales:

\begin{equation}
  S \left ( k \in \left [ 0.05 ; 0.5 \right ] \right ) \propto k^{-\alpha} \, .
  \label{eq:powerlaw}
\end{equation}

\begin{figure}
  \centering
  \includegraphics[width=\linewidth]{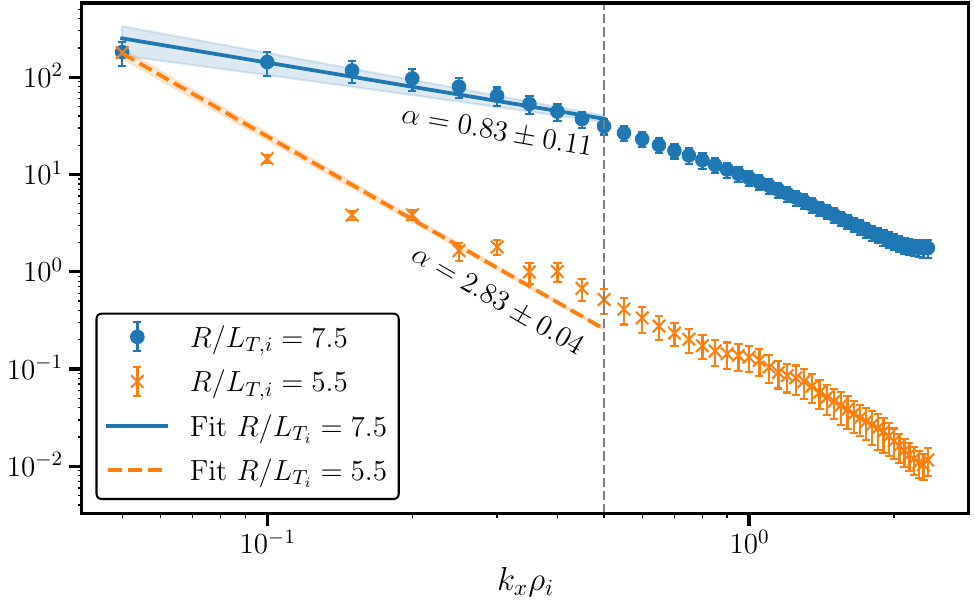}
  \caption{Chained simulations free energy $k_x$ spectra associated to turbulent ($R/L_{T,i}=7.5$, blue disks) and zonal ($R/L_{T,i} = 5.5$, orange crosses) regimes, and superimposed results of power law fits at large scales $k_x\rho_i \in [0.05,0.5]$, with exponents $\alpha$.}
  \label{fig:Fig_14}
\end{figure}

Figure~\ref{fig:Fig_14} shows the fitted $k_x$ spectra of free energy obtained from the chained simulations in the turbulent regime at ($R/L_{T,i}=7.5$, circles and blue solid line)
and the zonal flow dominated regime at ($R/L_{T,i}=5.5$, $\times$ markers and orange dashed line).
Only the ramp-down is considered as the ramp-up is basically the same in this extreme values.
The spectra are fitted using a power law as in \eqref{eq:powerlaw}.  The large scale spectrum in the zonal regime is generally steeper than the turbulent regime, with the fitted values of the exponent $\alpha$, such that
$\alpha(R/L_{T,i}=7.5) = 0.83$ vs. $\alpha(R/L_{T,i}=5.5) =2.83$.
This difference is a consequence of the role played by zonal flows in regulating the system. When zonal flows are sufficiently strong,
they scatter the turbulent fluctuations to smaller scales where they can be dissipated, resulting in a state with high free energy at large scale zonal modes and a rapid depletion of energy at higher $k_x$.

\begin{figure}[H]
  \centering
  \includegraphics[width=\linewidth]{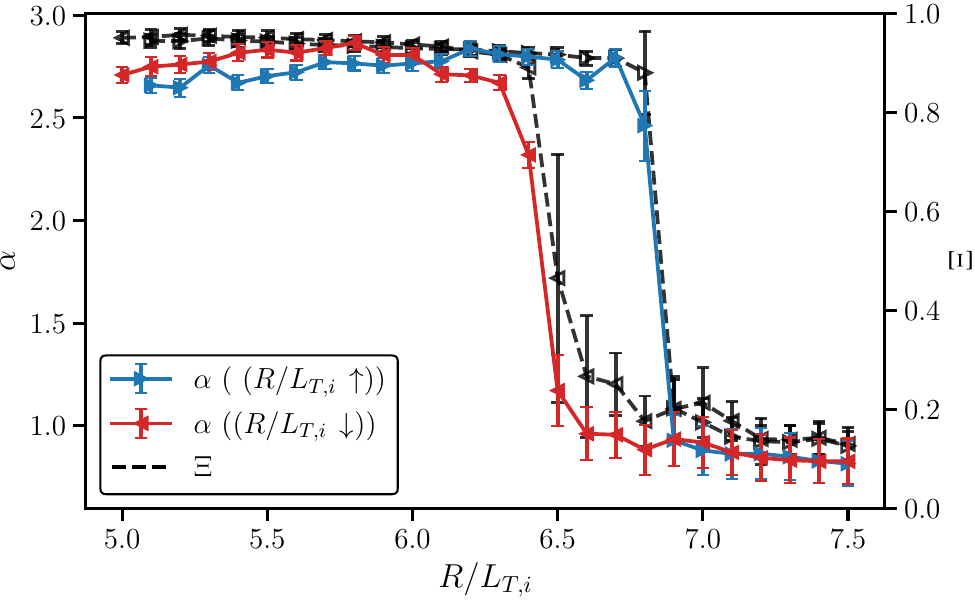}
  \caption{Evolution of the large scale slope $\alpha$ as a function of $R/L_{T,i}$, while decreasing (left-pointing red triangles) and increasing (right-pointing blue triangle) the temperature gradient. The zonal free energy fraction is also shown for comparison (black triangles and dashed line).}
  \label{fig:Fig_15}
\end{figure}

We can also consider the evolution of the spectral slope parameter $\alpha$ during the temperature gradient cycle of the chained simulations,
shown in Figure~\ref{fig:Fig_15}. As before, left-pointing triangle markers indicate the ramp-down, while right-pointing triangle markers denote the ramp-up of the temperature gradient.
We observe that the spectral slope exhibits two distinct plateaus with low values of the exponent, $\alpha \simeq 0.8-1$, in the turbulent regime, and high values, $\alpha \simeq 2.7-2.8$, in the zonal regime.

Large variations of $\alpha$ can be seen in the vicinity of the transition, where it increases from $\alpha \simeq 1$ to $\alpha \simeq 2.6$ as $R/L_{T,i}$ is reduced from $6.6$ to $6.4$.
During the ramp up, the transition occurs around $R/L_{T,i}=6.8\rightarrow6.9$, and $\alpha$ decreases from $\alpha \simeq 2.5$ to $\alpha \simeq 0.8$.
Overall, these results indicate that an hysteresis cycle can also be retrieved using the large-scale spectral slope $\alpha$ as an indicator of the type of regime characterizing the system.

For comparison, the evolution of the control parameter $\Xi$ is shown in the background of Figure~\ref{fig:Fig_15} using unfilled left/right-oriented triangles that are connected with a black dashed line.
Note that, the strongest variations of $\alpha$ happen a bit later than that of $\Xi$ during the turbulent to zonal transition, whereas they coincide for the back transition from zonal to turbulent states.
This suggests that the reorganization of the energy spectrum takes some time after the zonal flow becomes well established, but it is instantaneous when the zonal flow breaks apart.

\section{Conclusion \label{sec:Conclusion}}

Considering free energy, which is the quadratic, nonlinearly conserved quantity of gyrokinetics, we suggest a re-interpretation of the Dimits shift as being analogous to a "phase transition" in the thermodynamical sense.
The free energy is injected into the system by the ion temperature gradient, while its state of organization can be characterized by an order parameter given by the zonal fraction $\Xi$,
i.e. the ratio between the zonal and total, free energies. In this view, the Dimits shift is a state of higher order, preferred by the system because it minimizes the free energy, or maximizes its entropy production.

Unsing the nonlinear gyrokinetic code GENE, selecting numerical parameters that minimize zonal flow damping,
and performing independent scans of the temperature gradient around the Dimits shift, a sharp transition is observed between a zonal dominated phase for lowest values of the temperature gradient,
and a turbulence dominated phase at higher gradients, accompanied by the usual signatures of the Dimits shift, namely very low,
highly intermittent heat flux and strong zonal ${\bf E} \times {\bf B}$ flows in the zonal phase, 
and larger heat flux and low levels of zonal flows in the turbulent phase.

Performing chained simulations where the temperature gradient is varied slowly from one restart to another we obtained the same two phases,
but a more robust zonal dominated state with a zonal fraction reaching values close to unity while approaching the linear stability threshold.
Furthermore, an hysteresis loop is observed in the zonal fraction as the temperature gradient is decreased and increased back.
This hysteresis is also visible in the spectral indexes associated with the power law fits of the large scale $k_x$ spectra.
A slight delay is found between the two observables, (i.e. zonal free energy fraction, vs. the spectral slope) 
for the transition into the Dimits regime, suggesting that the zonal energy rises first,
and the spectral shape is modified later.

The analysis of the $k_x$ spectra suggests that the zonal flow dominated regime in the chained simulations, may be a result of the dominance of the box size mode with $k_x^0 = 0.05$.
In contrast, in independent simulations, the zonal wave-number spectra are found to be more diverse.
In the turbulent phase, the $k_x$ spectra show roughly two spectral exponents for large and small scales, and is noisier in the case of independent simulations.

One should also mention that series of chained and independent simulations have been performed with higher spectral resolution but have not reproduced clear hysteresis,
which suggests that this phenomenon is very sensitive to the dissipation rate of the zonal structures, since in these high resolution runs,
the effective zonal dissipation has been {\it a posteriori} found higher than the standard resolution cases.
This suggests that the minimization of zonal flow damping is an important aspect of the behavior that we report in this paper.
This is probably a consequence of the fact that increased zonal flow damping competes against the key physics of the hysteresis, which  is that once the zonal flows are established,
the system requires latent heat to destroy them. Nevertheless, the phase transition between zonal dominated and turbulent regimes is recovered with higher resolution simulations,
which appears to be a robust feature, in contrast to the presence of an hysteresis.

Here, the focus was electrostatic ITG with adiabatic electrons in a flux tube geometry for simplicity.
Future studies could include kinetic electrons, trapped electron physics, electromagnetic fluctuations, as well as global and flux driven formulations.
Comparison of flux driven and fixed gradient formulations using a simpler fluid model of dissipative drift waves suggest that the consequence of a coexistence of two phases (i.e. hysteresis) in a fixed gradient formulation
is the profile stiffness in a flux driven one \citep{Guillon2025Pflare}. 
Instead of having two different possible values of heat flux for a given temperature gradient, depending on the history of the system, which manifests itself in different zonal flow levels,
in the fixed gradient case, the flux driven formulation would give the same gradient for two different levels of the imposed flux, with different zonal flow levels.

Notice also that, heat flux being multi-valued for a given gradient appears to be a problem for fixed gradient formulation, which only happens near marginality.
However, the existence of a nonlinear threshold, such as the Dimits shift makes the boundary of marginality less clear.
Furthermore, since complex systems that are driven by fluxes, such as tokamaks tend towards a state of self-organized-criticality, in practice the physical system is likely to be not far from such a nonlinear threshold.
One way to get around this problem, may be to consider a parameter representing the zonal flow level (i.e. the zonal fraction as we discuss here) as an observable that is systematically reported together with the heat flux. In this perspective, instead of flux being a multivalued function of the gradient, it is interpreted as a single valued function of two variables: the gradient and the zonal fraction. Such an approach can also help improve the quasi-linear approach used to evaluate heat and particle fluxes.

One should also mention that the notion of phase transition is related to standard thermodynamics, while the system of interest here is clearly out of equilibrium,
since the imposed background density and temperature gradients act as a source of free energy.
More deeply than this nomenclature issue, the existence of two regimes could be characterized by some entropy production principle: preliminary results tend to show a minimal dissipation,
hence a minimal entropy production principle while close to marginality, i.e. in the zonal regime, while the free energy dissipation is observed very larger in the turbulent regime,
which could be linked to a maximal entropy production principle for large temperature gradients.
A detailed analysis with respect to entropy production exceeds by far the scope of the present work and is let for future studies.

\begin{acknowledgments}
This work was granted access to the HPC resources of IDCS support unit from \'Ecole Polytechnique,
as well as was provided with computer and storage resources by GENCI at IDRIS thanks to the grant 2025-AD010514291R1 on the supercomputer Jean Zay's the V100 partition.
The authors gratefully acknowledge P.~L.~Guillon for useful discussions.
This work has been carried out within the framework of the EUROfusion Consortium,
funded by the European Union via the Euratom Research and Training Programme (Grant Agreement No. 101052200—EUROfusion) and within the framework of the French Research Federation for Fusion Studies.
\end{acknowledgments}


\begin{thebibliography}{38}%
\makeatletter
\providecommand \@ifxundefined [1]{%
 \@ifx{#1\undefined}
}%
\providecommand \@ifnum [1]{%
 \ifnum #1\expandafter \@firstoftwo
 \else \expandafter \@secondoftwo
 \fi
}%
\providecommand \@ifx [1]{%
 \ifx #1\expandafter \@firstoftwo
 \else \expandafter \@secondoftwo
 \fi
}%
\providecommand \natexlab [1]{#1}%
\providecommand \enquote  [1]{``#1''}%
\providecommand \bibnamefont  [1]{#1}%
\providecommand \bibfnamefont [1]{#1}%
\providecommand \citenamefont [1]{#1}%
\providecommand \href@noop [0]{\@secondoftwo}%
\providecommand \href [0]{\begingroup \@sanitize@url \@href}%
\providecommand \@href[1]{\@@startlink{#1}\@@href}%
\providecommand \@@href[1]{\endgroup#1\@@endlink}%
\providecommand \@sanitize@url [0]{\catcode `\\12\catcode `\$12\catcode `\&12\catcode `\#12\catcode `\^12\catcode `\_12\catcode `\%12\relax}%
\providecommand \@@startlink[1]{}%
\providecommand \@@endlink[0]{}%
\providecommand \url  [0]{\begingroup\@sanitize@url \@url }%
\providecommand \@url [1]{\endgroup\@href {#1}{\urlprefix }}%
\providecommand \urlprefix  [0]{URL }%
\providecommand \Eprint [0]{\href }%
\providecommand \doibase [0]{https://doi.org/}%
\providecommand \selectlanguage [0]{\@gobble}%
\providecommand \bibinfo  [0]{\@secondoftwo}%
\providecommand \bibfield  [0]{\@secondoftwo}%
\providecommand \translation [1]{[#1]}%
\providecommand \BibitemOpen [0]{}%
\providecommand \bibitemStop [0]{}%
\providecommand \bibitemNoStop [0]{.\EOS\space}%
\providecommand \EOS [0]{\spacefactor3000\relax}%
\providecommand \BibitemShut  [1]{\csname bibitem#1\endcsname}%
\let\auto@bib@innerbib\@empty
\bibitem [{\citenamefont {Groebner}\ \emph {et~al.}(1986)\citenamefont {Groebner}, \citenamefont {Pfeiffer}, \citenamefont {Blau}, \citenamefont {Burrell}, \citenamefont {Fairbanks}, \citenamefont {Seraydarian}, \citenamefont {John},\ and\ \citenamefont {Stockdale}}]{Groebner1986}%
  \BibitemOpen
  \bibfield  {author} {\bibinfo {author} {\bibfnamefont {R.~J.}\ \bibnamefont {Groebner}}, \bibinfo {author} {\bibfnamefont {W.}~\bibnamefont {Pfeiffer}}, \bibinfo {author} {\bibfnamefont {F.~P.}\ \bibnamefont {Blau}}, \bibinfo {author} {\bibfnamefont {K.~H.}\ \bibnamefont {Burrell}}, \bibinfo {author} {\bibfnamefont {E.~S.}\ \bibnamefont {Fairbanks}}, \bibinfo {author} {\bibfnamefont {R.~P.}\ \bibnamefont {Seraydarian}}, \bibinfo {author} {\bibfnamefont {H.}~\bibnamefont {John}},\ and\ \bibinfo {author} {\bibfnamefont {R.~E.}\ \bibnamefont {Stockdale}},\ }\bibfield  {title} {\bibinfo {title} {Experimentally inferred ion thermal diffusivity profiles in the doublet iii tokamak: Comparison with neoclassical theory},\ }\href {https://doi.org/10.1088/0029-5515/26/5/001} {\bibfield  {journal} {\bibinfo  {journal} {Nuclear Fusion}\ }\textbf {\bibinfo {volume} {26}},\ \bibinfo {pages} {543} (\bibinfo {year} {1986})}\BibitemShut {NoStop}%
\bibitem [{\citenamefont {Carreras}(1997)}]{Carreras1997}%
  \BibitemOpen
  \bibfield  {author} {\bibinfo {author} {\bibfnamefont {B.~A.}\ \bibnamefont {Carreras}},\ }\bibfield  {title} {\bibinfo {title} {Progress in anomalous transport research in toroidal magnetic confinement devices},\ }\href {https://doi.org/10.1109/27.650902} {\bibfield  {journal} {\bibinfo  {journal} {IEEE Transactions on Plasma Science}\ }\textbf {\bibinfo {volume} {25}},\ \bibinfo {pages} {1281} (\bibinfo {year} {1997})}\BibitemShut {NoStop}%
\bibitem [{\citenamefont {Wootton}\ \emph {et~al.}(1990)\citenamefont {Wootton}, \citenamefont {Carreras}, \citenamefont {Matsumoto}, \citenamefont {McGuire}, \citenamefont {Peebles}, \citenamefont {Ritz}, \citenamefont {Terry},\ and\ \citenamefont {Zweben}}]{Wootton1990}%
  \BibitemOpen
  \bibfield  {author} {\bibinfo {author} {\bibfnamefont {A.~J.}\ \bibnamefont {Wootton}}, \bibinfo {author} {\bibfnamefont {B.~A.}\ \bibnamefont {Carreras}}, \bibinfo {author} {\bibfnamefont {H.}~\bibnamefont {Matsumoto}}, \bibinfo {author} {\bibfnamefont {K.}~\bibnamefont {McGuire}}, \bibinfo {author} {\bibfnamefont {W.~A.}\ \bibnamefont {Peebles}}, \bibinfo {author} {\bibfnamefont {C.~P.}\ \bibnamefont {Ritz}}, \bibinfo {author} {\bibfnamefont {P.~W.}\ \bibnamefont {Terry}},\ and\ \bibinfo {author} {\bibfnamefont {S.~J.}\ \bibnamefont {Zweben}},\ }\bibfield  {title} {\bibinfo {title} {Fluctuations and anomalous transport in tokamaks},\ }\href {https://doi.org/10.1063/1.859358} {\bibfield  {journal} {\bibinfo  {journal} {Physics of Fluids B: Plasma Physics}\ }\textbf {\bibinfo {volume} {2}},\ \bibinfo {pages} {2879} (\bibinfo {year} {1990})}\BibitemShut {NoStop}%
\bibitem [{\citenamefont {Coppi}\ \emph {et~al.}(1967)\citenamefont {Coppi}, \citenamefont {Rosenbluth},\ and\ \citenamefont {Sagdeev}}]{Coppi1967}%
  \BibitemOpen
  \bibfield  {author} {\bibinfo {author} {\bibfnamefont {B.}~\bibnamefont {Coppi}}, \bibinfo {author} {\bibfnamefont {M.~N.}\ \bibnamefont {Rosenbluth}},\ and\ \bibinfo {author} {\bibfnamefont {R.~Z.}\ \bibnamefont {Sagdeev}},\ }\bibfield  {title} {\bibinfo {title} {Instabilities due to temperature gradients in complex magnetic field configurations},\ }\href {https://doi.org/10.1063/1.1762151} {\bibfield  {journal} {\bibinfo  {journal} {Physics of Fluids}\ }\textbf {\bibinfo {volume} {10}},\ \bibinfo {pages} {582} (\bibinfo {year} {1967})}\BibitemShut {NoStop}%
\bibitem [{\citenamefont {Dimits}\ \emph {et~al.}(2000)\citenamefont {Dimits}, \citenamefont {Bateman}, \citenamefont {Beer}, \citenamefont {Cohen}, \citenamefont {Dorland}, \citenamefont {Hammett}, \citenamefont {Kim}, \citenamefont {Kinsey}, \citenamefont {Kotschenreuther}, \citenamefont {Kritz}, \citenamefont {Lao}, \citenamefont {Mandrekas}, \citenamefont {Nevins}, \citenamefont {Parker}, \citenamefont {Redd}, \citenamefont {Shumaker}, \citenamefont {Sydora},\ and\ \citenamefont {Weiland}}]{Dimits2000}%
  \BibitemOpen
  \bibfield  {author} {\bibinfo {author} {\bibfnamefont {A.~M.}\ \bibnamefont {Dimits}}, \bibinfo {author} {\bibfnamefont {G.}~\bibnamefont {Bateman}}, \bibinfo {author} {\bibfnamefont {M.~A.}\ \bibnamefont {Beer}}, \bibinfo {author} {\bibfnamefont {B.~I.}\ \bibnamefont {Cohen}}, \bibinfo {author} {\bibfnamefont {W.}~\bibnamefont {Dorland}}, \bibinfo {author} {\bibfnamefont {G.~W.}\ \bibnamefont {Hammett}}, \bibinfo {author} {\bibfnamefont {C.}~\bibnamefont {Kim}}, \bibinfo {author} {\bibfnamefont {J.~E.}\ \bibnamefont {Kinsey}}, \bibinfo {author} {\bibfnamefont {M.}~\bibnamefont {Kotschenreuther}}, \bibinfo {author} {\bibfnamefont {A.~H.}\ \bibnamefont {Kritz}}, \bibinfo {author} {\bibfnamefont {L.~L.}\ \bibnamefont {Lao}}, \bibinfo {author} {\bibfnamefont {J.}~\bibnamefont {Mandrekas}}, \bibinfo {author} {\bibfnamefont {W.~M.}\ \bibnamefont {Nevins}}, \bibinfo {author} {\bibfnamefont {S.~E.}\ \bibnamefont {Parker}}, \bibinfo {author} {\bibfnamefont {A.~J.}\ \bibnamefont {Redd}}, \bibinfo {author} {\bibfnamefont {D.~E.}\ \bibnamefont {Shumaker}}, \bibinfo {author} {\bibfnamefont {R.}~\bibnamefont {Sydora}},\ and\ \bibinfo {author} {\bibfnamefont {J.}~\bibnamefont {Weiland}},\ }\bibfield  {title} {\bibinfo {title} {Comparisons and physics basis of tokamak transport models and turbulence simulations},\ }\href {https://doi.org/10.1063/1.873896} {\bibfield  {journal} {\bibinfo  {journal} {Physics of Plasmas}\ }\textbf {\bibinfo {volume} {7}},\ \bibinfo {pages} {969} (\bibinfo {year} {2000})}\BibitemShut {NoStop}%
\bibitem [{\citenamefont {Peeters}\ \emph {et~al.}(2016)\citenamefont {Peeters}, \citenamefont {Rath}, \citenamefont {Buchholz}, \citenamefont {Camenen}, \citenamefont {Candy}, \citenamefont {Casson}, \citenamefont {Grosshauser}, \citenamefont {Hornsby}, \citenamefont {Strintzi},\ and\ \citenamefont {Weikl}}]{Peeters2016}%
  \BibitemOpen
  \bibfield  {author} {\bibinfo {author} {\bibfnamefont {A.~G.}\ \bibnamefont {Peeters}}, \bibinfo {author} {\bibfnamefont {F.}~\bibnamefont {Rath}}, \bibinfo {author} {\bibfnamefont {R.}~\bibnamefont {Buchholz}}, \bibinfo {author} {\bibfnamefont {Y.}~\bibnamefont {Camenen}}, \bibinfo {author} {\bibfnamefont {J.}~\bibnamefont {Candy}}, \bibinfo {author} {\bibfnamefont {F.~J.}\ \bibnamefont {Casson}}, \bibinfo {author} {\bibfnamefont {S.~R.}\ \bibnamefont {Grosshauser}}, \bibinfo {author} {\bibfnamefont {W.~A.}\ \bibnamefont {Hornsby}}, \bibinfo {author} {\bibfnamefont {D.}~\bibnamefont {Strintzi}},\ and\ \bibinfo {author} {\bibfnamefont {A.}~\bibnamefont {Weikl}},\ }\bibfield  {title} {\bibinfo {title} {Gradient-driven flux-tube simulations of ion temperature gradient turbulence close to the non-linear threshold},\ }\href {https://doi.org/10.1063/1.4961231} {\bibfield  {journal} {\bibinfo  {journal} {Physics of Plasmas}\ }\textbf {\bibinfo {volume} {23}},\ \bibinfo {pages} {082517} (\bibinfo {year} {2016})}\BibitemShut {NoStop}%
\bibitem [{\citenamefont {Weikl}\ \emph {et~al.}(2017)\citenamefont {Weikl}, \citenamefont {Peeters}, \citenamefont {Rath}, \citenamefont {Grosshauser}, \citenamefont {Buchholz}, \citenamefont {Hornsby}, \citenamefont {Seiferling},\ and\ \citenamefont {Strintzi}}]{Weikl2017}%
  \BibitemOpen
  \bibfield  {author} {\bibinfo {author} {\bibfnamefont {A.}~\bibnamefont {Weikl}}, \bibinfo {author} {\bibfnamefont {A.~G.}\ \bibnamefont {Peeters}}, \bibinfo {author} {\bibfnamefont {F.}~\bibnamefont {Rath}}, \bibinfo {author} {\bibfnamefont {S.~R.}\ \bibnamefont {Grosshauser}}, \bibinfo {author} {\bibfnamefont {R.}~\bibnamefont {Buchholz}}, \bibinfo {author} {\bibfnamefont {W.~A.}\ \bibnamefont {Hornsby}}, \bibinfo {author} {\bibfnamefont {F.}~\bibnamefont {Seiferling}},\ and\ \bibinfo {author} {\bibfnamefont {D.}~\bibnamefont {Strintzi}},\ }\bibfield  {title} {\bibinfo {title} {Ion temperature gradient turbulence close to the finite heat flux threshold},\ }\href {https://doi.org/10.1063/1.4986035} {\bibfield  {journal} {\bibinfo  {journal} {Physics of Plasmas}\ }\textbf {\bibinfo {volume} {24}},\ \bibinfo {pages} {102317} (\bibinfo {year} {2017})}\BibitemShut {NoStop}%
\bibitem [{\citenamefont {Connor}\ and\ \citenamefont {Wilson}(2000)}]{Connor2000}%
  \BibitemOpen
  \bibfield  {author} {\bibinfo {author} {\bibfnamefont {J.~W.}\ \bibnamefont {Connor}}\ and\ \bibinfo {author} {\bibfnamefont {H.~R.}\ \bibnamefont {Wilson}},\ }\bibfield  {title} {\bibinfo {title} {A review of theories of the {L}-{H} transition},\ }\href {https://doi.org/10.1088/0741-3335/42/1/201} {\bibfield  {journal} {\bibinfo  {journal} {Plasma Physics and Controlled Fusion}\ }\textbf {\bibinfo {volume} {42}},\ \bibinfo {pages} {R1} (\bibinfo {year} {2000})}\BibitemShut {NoStop}%
\bibitem [{\citenamefont {Wagner}\ \emph {et~al.}(1982)\citenamefont {Wagner}, \citenamefont {Becker}, \citenamefont {Behringer}, \citenamefont {Campbell}, \citenamefont {Eberhagen}, \citenamefont {Engelhardt}, \citenamefont {Fussmann}, \citenamefont {Gehre}, \citenamefont {Gernhardt}, \citenamefont {von Gierke}, \citenamefont {Haas}, \citenamefont {Huang}, \citenamefont {Karger}, \citenamefont {Keilhacker}, \citenamefont {Kl{\"u}ber}, \citenamefont {Kornherr}, \citenamefont {Lackner}, \citenamefont {Lisitano}, \citenamefont {Lister}, \citenamefont {Mayer}, \citenamefont {Meisel}, \citenamefont {M{\"u}ller}, \citenamefont {Murmann}, \citenamefont {Niedermeyer}, \citenamefont {Poschenrieder}, \citenamefont {Rapp}, \citenamefont {R{\"o}hr}, \citenamefont {Schneider}, \citenamefont {Siller}, \citenamefont {Speth}, \citenamefont {St{\"a}bler}, \citenamefont {Steuer}, \citenamefont {Venus}, \citenamefont {Vollmer},\ and\ \citenamefont {Y{\"u}}}]{WagnerASDEXLH}%
  \BibitemOpen
  \bibfield  {author} {\bibinfo {author} {\bibfnamefont {F.}~\bibnamefont {Wagner}}, \bibinfo {author} {\bibfnamefont {G.}~\bibnamefont {Becker}}, \bibinfo {author} {\bibfnamefont {K.}~\bibnamefont {Behringer}}, \bibinfo {author} {\bibfnamefont {D.}~\bibnamefont {Campbell}}, \bibinfo {author} {\bibfnamefont {A.}~\bibnamefont {Eberhagen}}, \bibinfo {author} {\bibfnamefont {W.}~\bibnamefont {Engelhardt}}, \bibinfo {author} {\bibfnamefont {G.}~\bibnamefont {Fussmann}}, \bibinfo {author} {\bibfnamefont {O.}~\bibnamefont {Gehre}}, \bibinfo {author} {\bibfnamefont {J.}~\bibnamefont {Gernhardt}}, \bibinfo {author} {\bibfnamefont {G.}~\bibnamefont {von Gierke}}, \bibinfo {author} {\bibfnamefont {G.}~\bibnamefont {Haas}}, \bibinfo {author} {\bibfnamefont {M.}~\bibnamefont {Huang}}, \bibinfo {author} {\bibfnamefont {F.}~\bibnamefont {Karger}}, \bibinfo {author} {\bibfnamefont {M.}~\bibnamefont {Keilhacker}}, \bibinfo {author} {\bibfnamefont {O.}~\bibnamefont {Kl{\"u}ber}}, \bibinfo {author} {\bibfnamefont {M.}~\bibnamefont {Kornherr}}, \bibinfo {author} {\bibfnamefont {K.}~\bibnamefont {Lackner}}, \bibinfo {author} {\bibfnamefont {G.}~\bibnamefont {Lisitano}}, \bibinfo {author} {\bibfnamefont {G.~G.}\ \bibnamefont {Lister}}, \bibinfo {author} {\bibfnamefont {H.~M.}\ \bibnamefont {Mayer}}, \bibinfo {author} {\bibfnamefont {D.}~\bibnamefont {Meisel}}, \bibinfo {author} {\bibfnamefont {E.~R.}\ \bibnamefont {M{\"u}ller}}, \bibinfo {author} {\bibfnamefont {H.}~\bibnamefont {Murmann}}, \bibinfo {author} {\bibfnamefont {H.}~\bibnamefont {Niedermeyer}}, \bibinfo {author} {\bibfnamefont {W.}~\bibnamefont {Poschenrieder}}, \bibinfo {author} {\bibfnamefont {H.}~\bibnamefont {Rapp}}, \bibinfo {author} {\bibfnamefont {H.}~\bibnamefont {R{\"o}hr}}, \bibinfo {author} {\bibfnamefont {F.}~\bibnamefont {Schneider}}, \bibinfo {author} {\bibfnamefont {G.}~\bibnamefont {Siller}}, \bibinfo {author} {\bibfnamefont {E.}~\bibnamefont {Speth}}, \bibinfo {author} {\bibfnamefont {A.}~\bibnamefont {St{\"a}bler}}, \bibinfo {author} {\bibfnamefont {K.~H.}\ \bibnamefont {Steuer}}, \bibinfo {author} {\bibfnamefont {G.}~\bibnamefont {Venus}}, \bibinfo {author} {\bibfnamefont {O.}~\bibnamefont {Vollmer}},\ and\ \bibinfo {author} {\bibfnamefont {Z.}~\bibnamefont {Y{\"u}}},\ }\bibfield  {title} {\bibinfo {title} {Regime of improved confinement and high beta in neutral-beam-heated divertor discharges of the {ASDEX} tokamak},\ }\href {https://doi.org/10.1103/PhysRevLett.49.1408} {\bibfield  {journal} {\bibinfo  {journal} {Physical Review Letters}\ }\textbf {\bibinfo {volume} {49}},\ \bibinfo {pages} {1408} (\bibinfo {year} {1982})}\BibitemShut {NoStop}%
\bibitem [{\citenamefont {Burrell}\ \emph {et~al.}(1992)\citenamefont {Burrell}, \citenamefont {Carlstrom}, \citenamefont {Doyle}, \citenamefont {Finkenthal}, \citenamefont {Gohil}, \citenamefont {Groebner}, \citenamefont {Hillis}, \citenamefont {Kim}, \citenamefont {Matsumoto}, \citenamefont {Moyer}, \citenamefont {Osborne}, \citenamefont {Rettig}, \citenamefont {Peebles}, \citenamefont {Rhodes}, \citenamefont {John}, \citenamefont {Stambaugh}, \citenamefont {Wade},\ and\ \citenamefont {Watkins}}]{Burrell1992}%
  \BibitemOpen
  \bibfield  {author} {\bibinfo {author} {\bibfnamefont {K.~H.}\ \bibnamefont {Burrell}}, \bibinfo {author} {\bibfnamefont {T.~N.}\ \bibnamefont {Carlstrom}}, \bibinfo {author} {\bibfnamefont {E.~J.}\ \bibnamefont {Doyle}}, \bibinfo {author} {\bibfnamefont {D.}~\bibnamefont {Finkenthal}}, \bibinfo {author} {\bibfnamefont {P.}~\bibnamefont {Gohil}}, \bibinfo {author} {\bibfnamefont {R.~J.}\ \bibnamefont {Groebner}}, \bibinfo {author} {\bibfnamefont {D.~L.}\ \bibnamefont {Hillis}}, \bibinfo {author} {\bibfnamefont {J.}~\bibnamefont {Kim}}, \bibinfo {author} {\bibfnamefont {H.}~\bibnamefont {Matsumoto}}, \bibinfo {author} {\bibfnamefont {R.~A.}\ \bibnamefont {Moyer}}, \bibinfo {author} {\bibfnamefont {T.~H.}\ \bibnamefont {Osborne}}, \bibinfo {author} {\bibfnamefont {C.~L.}\ \bibnamefont {Rettig}}, \bibinfo {author} {\bibfnamefont {W.~A.}\ \bibnamefont {Peebles}}, \bibinfo {author} {\bibfnamefont {T.~L.}\ \bibnamefont {Rhodes}}, \bibinfo {author} {\bibfnamefont {H.~S.}\ \bibnamefont {John}}, \bibinfo {author} {\bibfnamefont {R.~D.}\ \bibnamefont {Stambaugh}}, \bibinfo {author} {\bibfnamefont {M.~R.}\ \bibnamefont {Wade}},\ and\ \bibinfo {author} {\bibfnamefont {J.~G.}\ \bibnamefont {Watkins}},\ }\bibfield  {title} {\bibinfo {title} {Physics of the {L}-mode to {H}-mode transition in tokamaks},\ }\href {https://doi.org/10.1088/0741-3335/34/13/014} {\bibfield  {journal} {\bibinfo  {journal} {Plasma Physics and Controlled Fusion}\ }\textbf {\bibinfo {volume} {34}},\ \bibinfo {pages} {1859} (\bibinfo {year} {1992})}\BibitemShut {NoStop}%
\bibitem [{\citenamefont {Malkov}\ and\ \citenamefont {Diamond}(2009)}]{Malkov2009}%
  \BibitemOpen
  \bibfield  {author} {\bibinfo {author} {\bibfnamefont {M.~A.}\ \bibnamefont {Malkov}}\ and\ \bibinfo {author} {\bibfnamefont {P.~H.}\ \bibnamefont {Diamond}},\ }\bibfield  {title} {\bibinfo {title} {Weak hysteresis in a simplified model of the {L--H} transition},\ }\href {https://doi.org/10.1063/1.3062834} {\bibfield  {journal} {\bibinfo  {journal} {Physics of Plasmas}\ }\textbf {\bibinfo {volume} {16}},\ \bibinfo {pages} {012504} (\bibinfo {year} {2009})}\BibitemShut {NoStop}%
\bibitem [{\citenamefont {Itoh}\ \emph {et~al.}(2017)\citenamefont {Itoh}, \citenamefont {Itoh}, \citenamefont {Ida}, \citenamefont {Inagaki}, \citenamefont {Kamada}, \citenamefont {Kamiya}, \citenamefont {Dong}, \citenamefont {Hidalgo}, \citenamefont {Evans}, \citenamefont {Ko}, \citenamefont {Park}, \citenamefont {Tokuzawa}, \citenamefont {Kubo}, \citenamefont {Kobayashi}, \citenamefont {Kosuga}, \citenamefont {Sasaki}, \citenamefont {Yun}, \citenamefont {Song}, \citenamefont {Kasuya}, \citenamefont {Nagashima}, \citenamefont {Moon}, \citenamefont {Yoshinuma}, \citenamefont {Makino}, \citenamefont {Tsujimura}, \citenamefont {Tsuchiya},\ and\ \citenamefont {Stroth}}]{Itoh2017}%
  \BibitemOpen
  \bibfield  {author} {\bibinfo {author} {\bibfnamefont {K.}~\bibnamefont {Itoh}}, \bibinfo {author} {\bibfnamefont {S.-I.}\ \bibnamefont {Itoh}}, \bibinfo {author} {\bibfnamefont {K.}~\bibnamefont {Ida}}, \bibinfo {author} {\bibfnamefont {S.}~\bibnamefont {Inagaki}}, \bibinfo {author} {\bibfnamefont {Y.}~\bibnamefont {Kamada}}, \bibinfo {author} {\bibfnamefont {K.}~\bibnamefont {Kamiya}}, \bibinfo {author} {\bibfnamefont {J.~Q.}\ \bibnamefont {Dong}}, \bibinfo {author} {\bibfnamefont {C.}~\bibnamefont {Hidalgo}}, \bibinfo {author} {\bibfnamefont {T.}~\bibnamefont {Evans}}, \bibinfo {author} {\bibfnamefont {W.~H.}\ \bibnamefont {Ko}}, \bibinfo {author} {\bibfnamefont {H.}~\bibnamefont {Park}}, \bibinfo {author} {\bibfnamefont {T.}~\bibnamefont {Tokuzawa}}, \bibinfo {author} {\bibfnamefont {S.}~\bibnamefont {Kubo}}, \bibinfo {author} {\bibfnamefont {T.}~\bibnamefont {Kobayashi}}, \bibinfo {author} {\bibfnamefont {Y.}~\bibnamefont {Kosuga}}, \bibinfo {author} {\bibfnamefont {M.}~\bibnamefont {Sasaki}}, \bibinfo {author} {\bibfnamefont {G.~S.}\ \bibnamefont {Yun}}, \bibinfo {author} {\bibfnamefont {S.~D.}\ \bibnamefont {Song}}, \bibinfo {author} {\bibfnamefont {N.}~\bibnamefont {Kasuya}}, \bibinfo {author} {\bibfnamefont {Y.}~\bibnamefont {Nagashima}}, \bibinfo {author} {\bibfnamefont {C.}~\bibnamefont {Moon}}, \bibinfo {author} {\bibfnamefont {M.}~\bibnamefont {Yoshinuma}}, \bibinfo {author} {\bibfnamefont {R.}~\bibnamefont {Makino}}, \bibinfo {author} {\bibfnamefont {T.}~\bibnamefont {Tsujimura}}, \bibinfo {author} {\bibfnamefont {H.}~\bibnamefont {Tsuchiya}},\ and\ \bibinfo {author} {\bibfnamefont {U.}~\bibnamefont {Stroth}},\ }\bibfield  {title} {\bibinfo {title} {Hysteresis and fast timescales in transport relations of toroidal plasmas},\ }\href {https://doi.org/10.1088/1741-4326/aa796a} {\bibfield  {journal} {\bibinfo  {journal} {Nuclear Fusion}\ }\textbf {\bibinfo {volume} {57}},\ \bibinfo {pages} {102021} (\bibinfo {year} {2017})}\BibitemShut {NoStop}%
\bibitem [{\citenamefont {Gravier}\ \emph {et~al.}(2017)\citenamefont {Gravier}, \citenamefont {Lesur}, \citenamefont {Reveille}, \citenamefont {Drouot},\ and\ \citenamefont {M{\'e}dina}}]{Gravier2017}%
  \BibitemOpen
  \bibfield  {author} {\bibinfo {author} {\bibfnamefont {E.}~\bibnamefont {Gravier}}, \bibinfo {author} {\bibfnamefont {M.}~\bibnamefont {Lesur}}, \bibinfo {author} {\bibfnamefont {T.}~\bibnamefont {Reveille}}, \bibinfo {author} {\bibfnamefont {T.}~\bibnamefont {Drouot}},\ and\ \bibinfo {author} {\bibfnamefont {J.}~\bibnamefont {M{\'e}dina}},\ }\bibfield  {title} {\bibinfo {title} {Transport hysteresis and zonal flow stimulation in magnetized plasmas},\ }\href {https://doi.org/10.1088/1741-4326/aa8c4c} {\bibfield  {journal} {\bibinfo  {journal} {Nuclear Fusion}\ }\textbf {\bibinfo {volume} {57}},\ \bibinfo {pages} {124001} (\bibinfo {year} {2017})}\BibitemShut {NoStop}%
\bibitem [{\citenamefont {Grander}\ \emph {et~al.}(2024)\citenamefont {Grander}, \citenamefont {Locker},\ and\ \citenamefont {Kendl}}]{Grander2024}%
  \BibitemOpen
  \bibfield  {author} {\bibinfo {author} {\bibfnamefont {F.}~\bibnamefont {Grander}}, \bibinfo {author} {\bibfnamefont {F.~F.}\ \bibnamefont {Locker}},\ and\ \bibinfo {author} {\bibfnamefont {A.}~\bibnamefont {Kendl}},\ }\bibfield  {title} {\bibinfo {title} {Hysteresis in the gyrofluid resistive drift wave turbulence to zonal flow transition},\ }\href {https://doi.org/10.1063/5.0202720} {\bibfield  {journal} {\bibinfo  {journal} {Physics of Plasmas}\ }\textbf {\bibinfo {volume} {31}},\ \bibinfo {pages} {052301} (\bibinfo {year} {2024})}\BibitemShut {NoStop}%
\bibitem [{\citenamefont {Guillon}\ and\ \citenamefont {G{\"u}rcan}(2025)}]{Guillon2025}%
  \BibitemOpen
  \bibfield  {author} {\bibinfo {author} {\bibfnamefont {P.~L.}\ \bibnamefont {Guillon}}\ and\ \bibinfo {author} {\bibfnamefont {{\"O}.~D.}\ \bibnamefont {G{\"u}rcan}},\ }\bibfield  {title} {\bibinfo {title} {Phase transition from turbulence to zonal flows in the {H}asegawa--{W}akatani system},\ }\href {https://doi.org/10.1063/5.0242282} {\bibfield  {journal} {\bibinfo  {journal} {Physics of Plasmas}\ }\textbf {\bibinfo {volume} {32}},\ \bibinfo {pages} {012306} (\bibinfo {year} {2025})}\BibitemShut {NoStop}%
\bibitem [{\citenamefont {Grander}\ \emph {et~al.}(2026)\citenamefont {Grander}, \citenamefont {Gröfler}, \citenamefont {Locker}, \citenamefont {Rinner},\ and\ \citenamefont {Kendl}}]{grander2026mergingzonalflowsgyrofluid}%
  \BibitemOpen
  \bibfield  {author} {\bibinfo {author} {\bibfnamefont {F.}~\bibnamefont {Grander}}, \bibinfo {author} {\bibfnamefont {T.}~\bibnamefont {Gröfler}}, \bibinfo {author} {\bibfnamefont {F.~F.}\ \bibnamefont {Locker}}, \bibinfo {author} {\bibfnamefont {M.}~\bibnamefont {Rinner}},\ and\ \bibinfo {author} {\bibfnamefont {A.}~\bibnamefont {Kendl}},\ }\href {https://arxiv.org/abs/2602.23007} {\bibinfo {title} {Merging of zonal flows in gyrofluid resistive drift-wave turbulence}} (\bibinfo {year} {2026}),\ \Eprint {https://arxiv.org/abs/2602.23007} {arXiv:2602.23007 [physics.plasm-ph]} \BibitemShut {NoStop}%
\bibitem [{\citenamefont {Jenko}\ \emph {et~al.}(2000)\citenamefont {Jenko}, \citenamefont {Dorland}, \citenamefont {Kotschenreuther},\ and\ \citenamefont {Rogers}}]{Jenko2000}%
  \BibitemOpen
  \bibfield  {author} {\bibinfo {author} {\bibfnamefont {F.}~\bibnamefont {Jenko}}, \bibinfo {author} {\bibfnamefont {W.}~\bibnamefont {Dorland}}, \bibinfo {author} {\bibfnamefont {M.}~\bibnamefont {Kotschenreuther}},\ and\ \bibinfo {author} {\bibfnamefont {B.~N.}\ \bibnamefont {Rogers}},\ }\bibfield  {title} {\bibinfo {title} {Electron temperature gradient driven turbulence},\ }\href {https://doi.org/10.1063/1.874014} {\bibfield  {journal} {\bibinfo  {journal} {Physics of Plasmas}\ }\textbf {\bibinfo {volume} {7}},\ \bibinfo {pages} {1904} (\bibinfo {year} {2000})}\BibitemShut {NoStop}%
\bibitem [{\citenamefont {Brizard}\ and\ \citenamefont {Hahm}(2007)}]{Brizard2007}%
  \BibitemOpen
  \bibfield  {author} {\bibinfo {author} {\bibfnamefont {A.~J.}\ \bibnamefont {Brizard}}\ and\ \bibinfo {author} {\bibfnamefont {T.~S.}\ \bibnamefont {Hahm}},\ }\bibfield  {title} {\bibinfo {title} {Foundations of nonlinear gyrokinetic theory},\ }\href {https://doi.org/10.1103/RevModPhys.79.421} {\bibfield  {journal} {\bibinfo  {journal} {Reviews of Modern Physics}\ }\textbf {\bibinfo {volume} {79}},\ \bibinfo {pages} {421} (\bibinfo {year} {2007})}\BibitemShut {NoStop}%
\bibitem [{\citenamefont {Beer}(1994)}]{Beer1994}%
  \BibitemOpen
  \bibfield  {author} {\bibinfo {author} {\bibfnamefont {M.~A.}\ \bibnamefont {Beer}},\ }\emph {\bibinfo {title} {Gyrofluid Models of Turbulent Transport in Tokamaks}},\ \href@noop {} {Ph.D. thesis},\ \bibinfo  {school} {Princeton University} (\bibinfo {year} {1994})\BibitemShut {NoStop}%
\bibitem [{\citenamefont {Candy}\ \emph {et~al.}(2006)\citenamefont {Candy}, \citenamefont {Waltz}, \citenamefont {Parker},\ and\ \citenamefont {Chen}}]{Candy2006}%
  \BibitemOpen
  \bibfield  {author} {\bibinfo {author} {\bibfnamefont {J.}~\bibnamefont {Candy}}, \bibinfo {author} {\bibfnamefont {R.~E.}\ \bibnamefont {Waltz}}, \bibinfo {author} {\bibfnamefont {S.~E.}\ \bibnamefont {Parker}},\ and\ \bibinfo {author} {\bibfnamefont {Y.}~\bibnamefont {Chen}},\ }\bibfield  {title} {\bibinfo {title} {Relevance of the parallel nonlinearity in gyrokinetic simulations of tokamak plasmas},\ }\href {https://doi.org/10.1063/1.2220536} {\bibfield  {journal} {\bibinfo  {journal} {Physics of Plasmas}\ }\textbf {\bibinfo {volume} {13}},\ \bibinfo {pages} {074501} (\bibinfo {year} {2006})}\BibitemShut {NoStop}%
\bibitem [{\citenamefont {Morel}\ \emph {et~al.}(2011)\citenamefont {Morel}, \citenamefont {Navarro}, \citenamefont {Albrecht-Marc}, \citenamefont {Carati}, \citenamefont {Merz}, \citenamefont {G{\"o}rler},\ and\ \citenamefont {Jenko}}]{Morel2011}%
  \BibitemOpen
  \bibfield  {author} {\bibinfo {author} {\bibfnamefont {P.}~\bibnamefont {Morel}}, \bibinfo {author} {\bibfnamefont {A.~B.}\ \bibnamefont {Navarro}}, \bibinfo {author} {\bibfnamefont {M.}~\bibnamefont {Albrecht-Marc}}, \bibinfo {author} {\bibfnamefont {D.}~\bibnamefont {Carati}}, \bibinfo {author} {\bibfnamefont {F.}~\bibnamefont {Merz}}, \bibinfo {author} {\bibfnamefont {T.}~\bibnamefont {G{\"o}rler}},\ and\ \bibinfo {author} {\bibfnamefont {F.}~\bibnamefont {Jenko}},\ }\bibfield  {title} {\bibinfo {title} {Gyrokinetic large eddy simulations},\ }\href {https://doi.org/10.1063/1.3601053} {\bibfield  {journal} {\bibinfo  {journal} {Physics of Plasmas}\ }\textbf {\bibinfo {volume} {18}},\ \bibinfo {pages} {072301} (\bibinfo {year} {2011})}\BibitemShut {NoStop}%
\bibitem [{\citenamefont {Morel}\ \emph {et~al.}(2012)\citenamefont {Morel}, \citenamefont {Navarro}, \citenamefont {Albrecht-Marc}, \citenamefont {Carati}, \citenamefont {Merz}, \citenamefont {G{\"o}rler},\ and\ \citenamefont {Jenko}}]{Morel2012}%
  \BibitemOpen
  \bibfield  {author} {\bibinfo {author} {\bibfnamefont {P.}~\bibnamefont {Morel}}, \bibinfo {author} {\bibfnamefont {A.~B.}\ \bibnamefont {Navarro}}, \bibinfo {author} {\bibfnamefont {M.}~\bibnamefont {Albrecht-Marc}}, \bibinfo {author} {\bibfnamefont {D.}~\bibnamefont {Carati}}, \bibinfo {author} {\bibfnamefont {F.}~\bibnamefont {Merz}}, \bibinfo {author} {\bibfnamefont {T.}~\bibnamefont {G{\"o}rler}},\ and\ \bibinfo {author} {\bibfnamefont {F.}~\bibnamefont {Jenko}},\ }\bibfield  {title} {\bibinfo {title} {Dynamic procedure for filtered gyrokinetic simulations},\ }\href {https://doi.org/10.1063/1.3677366} {\bibfield  {journal} {\bibinfo  {journal} {Physics of Plasmas}\ }\textbf {\bibinfo {volume} {19}},\ \bibinfo {pages} {012311} (\bibinfo {year} {2012})}\BibitemShut {NoStop}%
\bibitem [{\citenamefont {Romanelli}(1989)}]{Romanelli1989}%
  \BibitemOpen
  \bibfield  {author} {\bibinfo {author} {\bibfnamefont {F.}~\bibnamefont {Romanelli}},\ }\bibfield  {title} {\bibinfo {title} {Ion temperature-gradient-driven modes and anomalous ion transport in tokamaks},\ }\href {https://doi.org/10.1063/1.859023} {\bibfield  {journal} {\bibinfo  {journal} {Physics of Fluids B: Plasma Physics}\ }\textbf {\bibinfo {volume} {1}},\ \bibinfo {pages} {1018} (\bibinfo {year} {1989})}\BibitemShut {NoStop}%
\bibitem [{\citenamefont {Diamond}\ \emph {et~al.}(2005)\citenamefont {Diamond}, \citenamefont {Itoh}, \citenamefont {Itoh},\ and\ \citenamefont {Hahm}}]{Diamond2005Review}%
  \BibitemOpen
  \bibfield  {author} {\bibinfo {author} {\bibfnamefont {P.~H.}\ \bibnamefont {Diamond}}, \bibinfo {author} {\bibfnamefont {S.-I.}\ \bibnamefont {Itoh}}, \bibinfo {author} {\bibfnamefont {K.}~\bibnamefont {Itoh}},\ and\ \bibinfo {author} {\bibfnamefont {T.~S.}\ \bibnamefont {Hahm}},\ }\bibfield  {title} {\bibinfo {title} {Zonal flows in plasma—a review},\ }\href {https://doi.org/10.1088/0741-3335/47/5/R01} {\bibfield  {journal} {\bibinfo  {journal} {Plasma Physics and Controlled Fusion}\ }\textbf {\bibinfo {volume} {47}},\ \bibinfo {pages} {R35} (\bibinfo {year} {2005})}\BibitemShut {NoStop}%
\bibitem [{\citenamefont {G{\"u}rcan}\ and\ \citenamefont {Diamond}(2015)}]{Gurcan2015}%
  \BibitemOpen
  \bibfield  {author} {\bibinfo {author} {\bibfnamefont {{\"O}.~D.}\ \bibnamefont {G{\"u}rcan}}\ and\ \bibinfo {author} {\bibfnamefont {P.~H.}\ \bibnamefont {Diamond}},\ }\bibfield  {title} {\bibinfo {title} {Zonal flows and pattern formation},\ }\href {https://doi.org/10.1088/1751-8113/48/29/293001} {\bibfield  {journal} {\bibinfo  {journal} {Journal of Physics A: Mathematical and Theoretical}\ }\textbf {\bibinfo {volume} {48}},\ \bibinfo {pages} {293001} (\bibinfo {year} {2015})}\BibitemShut {NoStop}%
\bibitem [{\citenamefont {Biglari}\ \emph {et~al.}(1990)\citenamefont {Biglari}, \citenamefont {Diamond},\ and\ \citenamefont {Terry}}]{Biglari1990}%
  \BibitemOpen
  \bibfield  {author} {\bibinfo {author} {\bibfnamefont {H.}~\bibnamefont {Biglari}}, \bibinfo {author} {\bibfnamefont {P.~H.}\ \bibnamefont {Diamond}},\ and\ \bibinfo {author} {\bibfnamefont {P.~W.}\ \bibnamefont {Terry}},\ }\bibfield  {title} {\bibinfo {title} {Influence of sheared poloidal rotation on edge turbulence},\ }\href {https://doi.org/10.1063/1.859529} {\bibfield  {journal} {\bibinfo  {journal} {Physics of Fluids B: Plasma Physics}\ }\textbf {\bibinfo {volume} {2}},\ \bibinfo {pages} {1} (\bibinfo {year} {1990})}\BibitemShut {NoStop}%
\bibitem [{\citenamefont {Malkov}\ \emph {et~al.}(2001)\citenamefont {Malkov}, \citenamefont {Diamond},\ and\ \citenamefont {Rosenbluth}}]{Malkov2001}%
  \BibitemOpen
  \bibfield  {author} {\bibinfo {author} {\bibfnamefont {M.~A.}\ \bibnamefont {Malkov}}, \bibinfo {author} {\bibfnamefont {P.~H.}\ \bibnamefont {Diamond}},\ and\ \bibinfo {author} {\bibfnamefont {M.~N.}\ \bibnamefont {Rosenbluth}},\ }\bibfield  {title} {\bibinfo {title} {On the nature of bursting in transport and turbulence in drift wave–zonal flow systems},\ }\href {https://doi.org/10.1063/1.1415424} {\bibfield  {journal} {\bibinfo  {journal} {Physics of Plasmas}\ }\textbf {\bibinfo {volume} {8}},\ \bibinfo {pages} {5073} (\bibinfo {year} {2001})}\BibitemShut {NoStop}%
\bibitem [{\citenamefont {Schmitz}\ \emph {et~al.}(2012)\citenamefont {Schmitz}, \citenamefont {Zeng}, \citenamefont {Rhodes}, \citenamefont {Hillesheim}, \citenamefont {Doyle}, \citenamefont {Groebner}, \citenamefont {Peebles}, \citenamefont {Burrell},\ and\ \citenamefont {Wang}}]{Schmitz2012}%
  \BibitemOpen
  \bibfield  {author} {\bibinfo {author} {\bibfnamefont {L.}~\bibnamefont {Schmitz}}, \bibinfo {author} {\bibfnamefont {L.}~\bibnamefont {Zeng}}, \bibinfo {author} {\bibfnamefont {T.~L.}\ \bibnamefont {Rhodes}}, \bibinfo {author} {\bibfnamefont {J.~C.}\ \bibnamefont {Hillesheim}}, \bibinfo {author} {\bibfnamefont {E.~J.}\ \bibnamefont {Doyle}}, \bibinfo {author} {\bibfnamefont {R.~J.}\ \bibnamefont {Groebner}}, \bibinfo {author} {\bibfnamefont {W.~A.}\ \bibnamefont {Peebles}}, \bibinfo {author} {\bibfnamefont {K.~H.}\ \bibnamefont {Burrell}},\ and\ \bibinfo {author} {\bibfnamefont {G.}~\bibnamefont {Wang}},\ }\bibfield  {title} {\bibinfo {title} {Role of zonal flow predator-prey oscillations in triggering the transition to {H}-mode confinement},\ }\href {https://doi.org/10.1103/PhysRevLett.108.155002} {\bibfield  {journal} {\bibinfo  {journal} {Physical Review Letters}\ }\textbf {\bibinfo {volume} {108}},\ \bibinfo {pages} {155002} (\bibinfo {year} {2012})}\BibitemShut {NoStop}%
\bibitem [{\citenamefont {Morel}\ \emph {et~al.}(2014)\citenamefont {Morel}, \citenamefont {G{\"u}rcan},\ and\ \citenamefont {Berionni}}]{Morel2014}%
  \BibitemOpen
  \bibfield  {author} {\bibinfo {author} {\bibfnamefont {P.}~\bibnamefont {Morel}}, \bibinfo {author} {\bibfnamefont {{\"O}.~D.}\ \bibnamefont {G{\"u}rcan}},\ and\ \bibinfo {author} {\bibfnamefont {V.}~\bibnamefont {Berionni}},\ }\bibfield  {title} {\bibinfo {title} {Characterization of predator--prey dynamics using the evolution of free energy in plasma turbulence},\ }\href {https://doi.org/10.1088/0741-3335/56/1/015002} {\bibfield  {journal} {\bibinfo  {journal} {Plasma Physics and Controlled Fusion}\ }\textbf {\bibinfo {volume} {56}},\ \bibinfo {pages} {015002} (\bibinfo {year} {2014})}\BibitemShut {NoStop}%
\bibitem [{\citenamefont {Kobayashi}\ \emph {et~al.}(2015)\citenamefont {Kobayashi}, \citenamefont {G{\"u}rcan},\ and\ \citenamefont {Diamond}}]{Kobayashi2015}%
  \BibitemOpen
  \bibfield  {author} {\bibinfo {author} {\bibfnamefont {S.}~\bibnamefont {Kobayashi}}, \bibinfo {author} {\bibfnamefont {{\"O}.~D.}\ \bibnamefont {G{\"u}rcan}},\ and\ \bibinfo {author} {\bibfnamefont {P.~H.}\ \bibnamefont {Diamond}},\ }\bibfield  {title} {\bibinfo {title} {Direct identification of predator-prey dynamics in gyrokinetic simulations},\ }\href {https://doi.org/10.1063/1.4930127} {\bibfield  {journal} {\bibinfo  {journal} {Physics of Plasmas}\ }\textbf {\bibinfo {volume} {22}},\ \bibinfo {pages} {090702} (\bibinfo {year} {2015})}\BibitemShut {NoStop}%
\bibitem [{\citenamefont {Lin}\ \emph {et~al.}(1999)\citenamefont {Lin}, \citenamefont {Hahm}, \citenamefont {Lee}, \citenamefont {Tang},\ and\ \citenamefont {Diamond}}]{Lin1999}%
  \BibitemOpen
  \bibfield  {author} {\bibinfo {author} {\bibfnamefont {Z.}~\bibnamefont {Lin}}, \bibinfo {author} {\bibfnamefont {T.~S.}\ \bibnamefont {Hahm}}, \bibinfo {author} {\bibfnamefont {W.~W.}\ \bibnamefont {Lee}}, \bibinfo {author} {\bibfnamefont {W.~M.}\ \bibnamefont {Tang}},\ and\ \bibinfo {author} {\bibfnamefont {P.~H.}\ \bibnamefont {Diamond}},\ }\bibfield  {title} {\bibinfo {title} {Effects of collisional zonal flow damping on turbulent transport},\ }\href {https://doi.org/10.1103/PhysRevLett.83.3645} {\bibfield  {journal} {\bibinfo  {journal} {Physical Review Letters}\ }\textbf {\bibinfo {volume} {83}},\ \bibinfo {pages} {3645} (\bibinfo {year} {1999})}\BibitemShut {NoStop}%
\bibitem [{\citenamefont {Rosenbluth}\ and\ \citenamefont {Hinton}(1998)}]{Rosenbluth1998}%
  \BibitemOpen
  \bibfield  {author} {\bibinfo {author} {\bibfnamefont {M.~N.}\ \bibnamefont {Rosenbluth}}\ and\ \bibinfo {author} {\bibfnamefont {F.~L.}\ \bibnamefont {Hinton}},\ }\bibfield  {title} {\bibinfo {title} {Poloidal flow driven by ion-temperature-gradient turbulence in tokamaks},\ }\href {https://doi.org/10.1103/PhysRevLett.80.724} {\bibfield  {journal} {\bibinfo  {journal} {Physical Review Letters}\ }\textbf {\bibinfo {volume} {80}},\ \bibinfo {pages} {724} (\bibinfo {year} {1998})}\BibitemShut {NoStop}%
\bibitem [{\citenamefont {Pueschel}\ \emph {et~al.}(2010)\citenamefont {Pueschel}, \citenamefont {Dannert},\ and\ \citenamefont {Jenko}}]{Pueschel2010}%
  \BibitemOpen
  \bibfield  {author} {\bibinfo {author} {\bibfnamefont {M.~J.}\ \bibnamefont {Pueschel}}, \bibinfo {author} {\bibfnamefont {T.}~\bibnamefont {Dannert}},\ and\ \bibinfo {author} {\bibfnamefont {F.}~\bibnamefont {Jenko}},\ }\bibfield  {title} {\bibinfo {title} {On the role of numerical dissipation in gyrokinetic vlasov simulations of plasma microturbulence},\ }\href {https://doi.org/10.1016/j.cpc.2010.04.010} {\bibfield  {journal} {\bibinfo  {journal} {Computer Physics Communications}\ }\textbf {\bibinfo {volume} {181}},\ \bibinfo {pages} {1428} (\bibinfo {year} {2010})}\BibitemShut {NoStop}%
\bibitem [{\citenamefont {Schekochihin}\ \emph {et~al.}(2008)\citenamefont {Schekochihin}, \citenamefont {Cowley}, \citenamefont {Dorland}, \citenamefont {Hammett}, \citenamefont {Howes}, \citenamefont {Plunk}, \citenamefont {Quataert},\ and\ \citenamefont {Tatsuno}}]{Schekochihin2008}%
  \BibitemOpen
  \bibfield  {author} {\bibinfo {author} {\bibfnamefont {A.~A.}\ \bibnamefont {Schekochihin}}, \bibinfo {author} {\bibfnamefont {S.~C.}\ \bibnamefont {Cowley}}, \bibinfo {author} {\bibfnamefont {W.}~\bibnamefont {Dorland}}, \bibinfo {author} {\bibfnamefont {G.~W.}\ \bibnamefont {Hammett}}, \bibinfo {author} {\bibfnamefont {G.~G.}\ \bibnamefont {Howes}}, \bibinfo {author} {\bibfnamefont {G.~G.}\ \bibnamefont {Plunk}}, \bibinfo {author} {\bibfnamefont {E.}~\bibnamefont {Quataert}},\ and\ \bibinfo {author} {\bibfnamefont {T.}~\bibnamefont {Tatsuno}},\ }\bibfield  {title} {\bibinfo {title} {Gyrokinetic turbulence: a nonlinear route to dissipation through phase space},\ }\href {https://doi.org/10.1088/0741-3335/50/12/124024} {\bibfield  {journal} {\bibinfo  {journal} {Plasma Physics and Controlled Fusion}\ }\textbf {\bibinfo {volume} {50}},\ \bibinfo {pages} {124024} (\bibinfo {year} {2008})}\BibitemShut {NoStop}%
\bibitem [{\citenamefont {Navarro}\ \emph {et~al.}(2011)\citenamefont {Navarro}, \citenamefont {Morel}, \citenamefont {Albrecht-Marc}, \citenamefont {Carati}, \citenamefont {Merz}, \citenamefont {G{\"o}rler},\ and\ \citenamefont {Jenko}}]{Banon2011}%
  \BibitemOpen
  \bibfield  {author} {\bibinfo {author} {\bibfnamefont {A.~B.}\ \bibnamefont {Navarro}}, \bibinfo {author} {\bibfnamefont {P.}~\bibnamefont {Morel}}, \bibinfo {author} {\bibfnamefont {M.}~\bibnamefont {Albrecht-Marc}}, \bibinfo {author} {\bibfnamefont {D.}~\bibnamefont {Carati}}, \bibinfo {author} {\bibfnamefont {F.}~\bibnamefont {Merz}}, \bibinfo {author} {\bibfnamefont {T.}~\bibnamefont {G{\"o}rler}},\ and\ \bibinfo {author} {\bibfnamefont {F.}~\bibnamefont {Jenko}},\ }\bibfield  {title} {\bibinfo {title} {Free energy balance in gyrokinetic turbulence},\ }\href {https://doi.org/10.1063/1.3632077} {\bibfield  {journal} {\bibinfo  {journal} {Physics of Plasmas}\ }\textbf {\bibinfo {volume} {18}},\ \bibinfo {pages} {092307} (\bibinfo {year} {2011})}\BibitemShut {NoStop}%
\bibitem [{\citenamefont {Dif-Pradalier}\ \emph {et~al.}(2015)\citenamefont {Dif-Pradalier}, \citenamefont {Hornung}, \citenamefont {Ghendrih}, \citenamefont {Sarazin}, \citenamefont {Clairet}, \citenamefont {Vermare}, \citenamefont {Diamond}, \citenamefont {Abiteboul}, \citenamefont {Cartier-Michaud}, \citenamefont {Ehrlacher}, \citenamefont {Est{\`e}ve}, \citenamefont {Garbet}, \citenamefont {Grandgirard}, \citenamefont {G{\"u}rcan}, \citenamefont {Hennequin}, \citenamefont {Kosuga}, \citenamefont {Latu}, \citenamefont {Maget}, \citenamefont {Morel}, \citenamefont {Norscini}, \citenamefont {Sabot},\ and\ \citenamefont {Storelli}}]{DifPradalier2015}%
  \BibitemOpen
  \bibfield  {author} {\bibinfo {author} {\bibfnamefont {G.}~\bibnamefont {Dif-Pradalier}}, \bibinfo {author} {\bibfnamefont {G.}~\bibnamefont {Hornung}}, \bibinfo {author} {\bibfnamefont {P.}~\bibnamefont {Ghendrih}}, \bibinfo {author} {\bibfnamefont {Y.}~\bibnamefont {Sarazin}}, \bibinfo {author} {\bibfnamefont {F.}~\bibnamefont {Clairet}}, \bibinfo {author} {\bibfnamefont {L.}~\bibnamefont {Vermare}}, \bibinfo {author} {\bibfnamefont {P.~H.}\ \bibnamefont {Diamond}}, \bibinfo {author} {\bibfnamefont {J.}~\bibnamefont {Abiteboul}}, \bibinfo {author} {\bibfnamefont {T.}~\bibnamefont {Cartier-Michaud}}, \bibinfo {author} {\bibfnamefont {C.}~\bibnamefont {Ehrlacher}}, \bibinfo {author} {\bibfnamefont {D.}~\bibnamefont {Est{\`e}ve}}, \bibinfo {author} {\bibfnamefont {X.}~\bibnamefont {Garbet}}, \bibinfo {author} {\bibfnamefont {V.}~\bibnamefont {Grandgirard}}, \bibinfo {author} {\bibfnamefont {{\"O}.~D.}\ \bibnamefont {G{\"u}rcan}}, \bibinfo {author} {\bibfnamefont {P.}~\bibnamefont {Hennequin}}, \bibinfo {author} {\bibfnamefont {Y.}~\bibnamefont {Kosuga}}, \bibinfo {author} {\bibfnamefont {G.}~\bibnamefont {Latu}}, \bibinfo {author} {\bibfnamefont {P.}~\bibnamefont {Maget}}, \bibinfo {author} {\bibfnamefont {P.}~\bibnamefont {Morel}}, \bibinfo {author} {\bibfnamefont {C.}~\bibnamefont {Norscini}}, \bibinfo {author} {\bibfnamefont {R.}~\bibnamefont {Sabot}},\ and\ \bibinfo {author} {\bibfnamefont {A.}~\bibnamefont {Storelli}},\ }\bibfield  {title} {\bibinfo {title} {Finding the elusive {E}$\times${B} staircase in magnetized plasmas},\ }\href {https://doi.org/10.1103/PhysRevLett.114.085004} {\bibfield  {journal} {\bibinfo  {journal} {Physical Review Letters}\ }\textbf {\bibinfo {volume} {114}},\ \bibinfo {pages} {085004} (\bibinfo {year} {2015})}\BibitemShut {NoStop}%
\bibitem [{\citenamefont {Lippert}\ \emph {et~al.}(2023)\citenamefont {Lippert}, \citenamefont {Rath},\ and\ \citenamefont {Peeters}}]{Lippert2023}%
  \BibitemOpen
  \bibfield  {author} {\bibinfo {author} {\bibfnamefont {M.}~\bibnamefont {Lippert}}, \bibinfo {author} {\bibfnamefont {F.}~\bibnamefont {Rath}},\ and\ \bibinfo {author} {\bibfnamefont {A.~G.}\ \bibnamefont {Peeters}},\ }\bibfield  {title} {\bibinfo {title} {Size convergence of the {E}$\times${B} staircase pattern in flux-tube simulations of ion temperature gradient-driven turbulence},\ }\href {https://doi.org/10.1063/5.0153305} {\bibfield  {journal} {\bibinfo  {journal} {Physics of Plasmas}\ }\textbf {\bibinfo {volume} {30}},\ \bibinfo {pages} {074501} (\bibinfo {year} {2023})}\BibitemShut {NoStop}%
\bibitem [{\citenamefont {Guillon}\ \emph {et~al.}(2025)\citenamefont {Guillon}, \citenamefont {G{\"u}rcan}, \citenamefont {Dif-Pradalier}, \citenamefont {Sarazin},\ and\ \citenamefont {Fedorczak}}]{Guillon2025Pflare}%
  \BibitemOpen
  \bibfield  {author} {\bibinfo {author} {\bibfnamefont {P.~L.}\ \bibnamefont {Guillon}}, \bibinfo {author} {\bibfnamefont {{\"O}.~D.}\ \bibnamefont {G{\"u}rcan}}, \bibinfo {author} {\bibfnamefont {G.}~\bibnamefont {Dif-Pradalier}}, \bibinfo {author} {\bibfnamefont {Y.}~\bibnamefont {Sarazin}},\ and\ \bibinfo {author} {\bibfnamefont {N.}~\bibnamefont {Fedorczak}},\ }\bibfield  {title} {\bibinfo {title} {Flux-driven turbulent transport using penalisation in the hasegawa--wakatani system},\ }\href {https://doi.org/10.1017/S0022377825100895} {\bibfield  {journal} {\bibinfo  {journal} {Journal of Plasma Physics}\ }\textbf {\bibinfo {volume} {91}},\ \bibinfo {pages} {E145} (\bibinfo {year} {2025})}\BibitemShut {NoStop}%
\end{thebibliography}
\end{document}